# Thermo-mechanical finite element analysis of the AA5086 alloy under warm forming conditions

D.M. Neto[a,*], J.M.P. Martins[a,b], P.M. Cunha[a], J.L. Alves[c], M.C. Oliveira[a], H. Laurent[d], L.F. Menezes[a]

[a]CEMMPRE, Department of Mechanical Engineering, University of Coimbra, Polo II, Rua Luís Reis Santos, Pinhal de Marrocos, 3030-788 Coimbra, Portugal

[b]Department of Mechanical Engineering, Centre for Mechanical Technology and Automation, Grids, University of Aveiro, Campus Universitário de Santiago, 3810-193 Aveiro, Portugal

[c]CMEMS, Microelectromechanical Systems Research Unit, University of Minho, Campus de Azurém, 4800-058 Guimarães, Portugal

[d]Univ. Bretagne Sud, FRE CNRS 3744, IRDL, F-56100 Lorient, France

## Abstract

Warm forming processes have been successfully applied at laboratory level to overcome some important drawbacks of the Al−Mg alloys, such as poor formability and large springback. However, the numerical simulation of these processes requires the adoption of coupled thermo-mechanical finite element analysis, using temperature-dependent material models. The numerical description of the thermo-mechanical behaviour can require a large set of experimental tests. These experimental tests should be performed under conditions identical to the ones observed in the forming process. In this study, the warm deep drawing of a cylindrical cup is analysed, including the split-ring test to assess the temperature effect on the springback. Based on the analysis of the forming process conditions, the thermo-mechanical behaviour of the AA5086 aluminium alloy is described by a rate-independent thermo-elasto-plastic material model. The hardening law adopted is temperature-dependent while the yield function is temperature-independent. Nevertheless, the yield criterion parameters are selected based on the temperature of the heated tools. In fact, the model assumes that the temperature of the tools is uniform and constant, adopting a variable interfacial heat transfer coefficient. The accuracy of the proposed finite element model is assessed by comparing numerical and experimental results. The predicted punch force, thickness distribution and earing profile are in very good agreement with the experimental measurements, when the anisotropic behaviour of the blank is accurately described. However, this does not guarantee a correct springback prediction, which is strongly influenced by the elastic properties, namely the Young's modulus.

[*]Corresponding author: Tel.: +351239790700, Fax: +351239790701.
E-mail addresses: diogo.neto@dem.uc.pt (D.M. Neto), joao.martins52@ua.pt (J.M.P. Martins), patrick.cunha@uc.pt (P.M. Cunha), jlalves@dem.uminho.pt (J.L. Alves), marta.oliveira@dem.uc.pt (M.C. Oliveira), herve.laurent@univ-ubs.fr (H. Laurent), luis.menezes@dem.uc.pt (L.F. Menezes)

# 1. Introduction

Nowadays, one of the big challenges of the automotive industry is the reduction of the vehicles weight in order to reduce the fuel consumption and the $CO_2$ emission. In order to address this issue, the lightweight design plays an important role in the energy economy and environmental protection (Zhang et al., 2007). Therefore, the adoption of aluminium alloys is increasing in the automotive applications, replacing the conventional mild steels (Hirsch, 2011). However, this class of materials presents low formability (Kim et al., 2006) and high springback (Grèze et al., 2010) at room temperature, which are two of the main drawbacks arising in deep drawing processes. Furthermore, the Al−Mg alloys (5xxx series) are known to exhibit the Portevin-Le Chatelier effect, which is responsible for non-aesthetic stretcher lines on the sheet surfaces due to the localization of plastic deformation in bands (Coër et al., 2013; Yilmaz, 2011). Thus, these aluminium alloys are more used in the production of stiffeners for inner body panels.

Since the increase of the material temperature leads to a decrease of the flow stress and an increase of ductility, the warm forming of aluminium alloy sheets (below the recrystallization temperature) has become an interesting alternative to the conventional cold sheet metal forming (Toros et al., 2008). Indeed, the creation of a temperature gradient from the bottom to the flange (heated die and cooled punch) provides the best results in terms of formability for the warm deep drawing of a cylindrical cup (Palumbo et al., 2007). The temperature gradient between the bottom of the cup and the flange area determines an inverse yielding gradient (the region with the highest temperature has the lowest yielding stress) fundamental for the safety of the cup. In fact, it causes: (i) lower drawing force due to the high temperature in the blank flange and (ii) higher material strength in the cup bottom, particularly in the punch fillet radius region (Palumbo et al., 2007; Wilson, 1988). According to (Bolt et al., 2001), the maximum height of a AA5754-O aluminium rectangular cup can be increased 65% by using warm forming, with the die at 250ºC and a cooled punch. The three aluminium alloys studied by (Li and Ghosh, 2004) exhibit a significant improvement in their formability in the biaxial warm forming at temperatures ranging from 200ºC to 350ºC. The effect of temperature on the forming limit diagram was experimentally investigated by (Naka et al., 2001) for the 5083-O aluminium alloy, showing a substantial improvement for low values of plastic strain rate. Since the temperature rise also affects the stress state of the component, the springback effect is also reduced at warm forming conditions (Grèze et al., 2010; Kim and Koç, 2008).

The finite element simulation of warm forming processes plays an important role for understanding the complex deformation mechanisms, essential in the development and improvement of this technology. Nevertheless, the accuracy of the numerical solutions is strongly dependent on the models adopted in the numerical analysis (Abedrabbo et al., 2007; Berisha et al., 2010). The influence of the thermal field on the mechanical response and *vice-versa* must be considered in the warm forming process. In addition to the highly non-linear material behaviour and frictional contact conditions, the complex nature of this forming process also arises from the transient thermo-mechanical coupled response (Laurent et al., 2015). The thermal analysis requires improved knowledge concerning the heat transfer (conduction, convection and radiation) and the heat generated, either by plastic deformation or frictional sliding (J M P Martins et al., 2016).

Since the mechanical behaviour of the blank is temperature-dependent (Li and Ghosh, 2003), the numerical analysis comprises the solution of the transient heat conduction problem in addition to the elasto-plastic problem. Therefore, the interdependence of the thermal and mechanical solutions requires the adoption of a thermo-mechanical coupling algorithm (Vaz and Lange, 2016; Wriggers et al., 1992). Typically, two different solution strategies are used in the finite element simulation: monolithic and staggered. In the monolithic coupling procedure all fields are solved simultaneously using a single system of equations (involving both displacements and temperatures), which leads to an unconditionally stable solution for the problem. On the other hand, using the staggered coupling procedure, the mechanical and the thermal fields are partitioned into two systems of equations, which are solved sequentially (Erbts and Düster, 2012). The staggered approach is more flexible from the numerical implementation point of view and requires less computational resources. In fact, this approach allows to use different time scales and spatial discretizations for each sub-problem (Armero and Simo, 1992). The analysis of coupled thermo-mechanical problems using staggered algorithms can be performed using two distinct methodologies: the isothermal split (Argyris and Doltsinis, 1981) or the adiabatic split (Armero and Simo, 1992). Regarding the isothermal split, the mechanical problem is solved at constant temperature and the thermal problem is solved for a fixed configuration. In case of the adiabatic split, the mechanical problem is solved at constant entropy, while the thermal problem is solved for a fixed configuration. The main drawback of the isothermal split is the conditional stability, which arises when the thermos-elastic effects play an important role. However, this is not really an issue for the classical materials used in metal forming (Armero and Simo, 1993). Thus, this approach is the one adopted in the current work, resorting to an algorithm that performs the interchange of information between the thermal and the mechanical problem, both in the prediction and in the correction phases, to try to take advantage of automatic time-step control techniques (Martins et al., 2017).

Previous studies indicate that both the hardening and the anisotropic behaviour are influenced by the strain rate and the temperature. In addition to the flow stress decrease with the temperature rise, (Kabirian et al., 2014) reported a negative strain rate sensitivity of the AA5182-O aluminium alloy at room temperature, which changes to positive strain rate sensitivity for temperatures higher than 100ºC. The experimental study performed by (Coër et al., 2011) on the AA5754-O aluminium alloy shows that the anisotropy coefficients are rather constant from room temperature up to 200ºC (relative variation less than 8%). The effect of the strain rate on the anisotropic behaviour of the AA5182-O aluminium alloy was experimentally examined by (Rahmaan et al., 2016), showing that the anisotropy coefficients are relatively insensitive to the strain rate. Since the material parameters are calculated from experimental data at discrete temperatures, curve fitting is commonly used in the constitutive equations to obtain the anisotropy coefficients and the hardening parameters as a function of temperature (Abedrabbo et al., 2006). In this context, the model proposed by (Abedrabbo et al., 2007) for the finite element analysis of warm forming processes highlights the importance of using the thermal analysis in the description of the yield surface to model the warm forming process more accurately. Identical conclusion was found by (Kurukuri et al., 2009) in the deep drawing of a cylindrical cup under warm conditions. The influence of the yield locus shape on the predicted thickness distribution was also studied by (van den Boogaard and Huétink, 2006) using the deep drawing of an AA5754-O aluminium cylindrical cup.

The purpose of this study is to analyse numerically the warm forming conditions of the AA5086 alloy using thermo-mechanical finite element analysis. In order to assess the accuracy of the adopted model, the cylindrical cup proposed as benchmark at the conference Numisheet 2016 is the example selected (Manach et al., 2016). Furthermore, the influence of the process temperature conditions on the springback is evaluated through the split-ring test. The main process parameters studied are the punch force evolution, the thickness distribution measured in several directions, the earing profile as well as the ring opening. Section 2 contains a brief description of the warm forming process, including the springback evaluation using the split-ring test, as well as the experimental thermo-mechanical characterization of the AA5086 alloy. The temperature influence on the mechanical behaviour of this aluminium alloy is described by the thermo-elasto-plastic constitutive model presented in Section 3. The proposed finite element model is presented in Section 4, including the staggered coupling procedure, which is important for modelling of the thermo-mechanical contact conditions. Section 5 comprises the comparison between numerical and experimental results, which were kindly provided by the benchmark committee (Manach et al., 2016). The main conclusions of this study are discussed in Section 6.

## 2. Experimental procedure

This section contains a brief presentation of the warm forming process and the posterior springback evaluation using the split-ring test. The selected example was proposed in the conference Numisheet 2016, as a benchmark to evaluate the springback of an AA5086 alloy under warm forming conditions (Manach et al., 2016). In order to perform an accurate thermo-mechanical characterization of the Al−Mg alloy sheet, the benchmark committee also provided the data from the following tests: (i) uniaxial tensile tests performed under isothermal conditions at different temperatures and strain rates; (ii) monotonous and reverse shear tests at different temperatures and (iii) biaxial test (hydraulic bulge) at room temperature (Manach et al., 2016). Although many of the details included in this section can be found in (Manach et al., 2016), it is important to discuss them to fully understand the assumptions adopted in the numerical model.

### 2.1. Warm forming process

The warm deep drawing of a cylindrical cup is the example considered in this benchmark, which is based on the study of (Laurent et al., 2015). Figure 1 presents schematically the geometry of the forming tools (die, punch and blank-holder) used in the warm deep drawing process, which are made of hardened XC38CrMoV5 steel. The main dimensions of the tools (axisymmetric) are given in Table 1. The experimental cylindrical cup forming tests were carried out on a Zwick/Roell BUP200 machine (Manach et al., 2016).

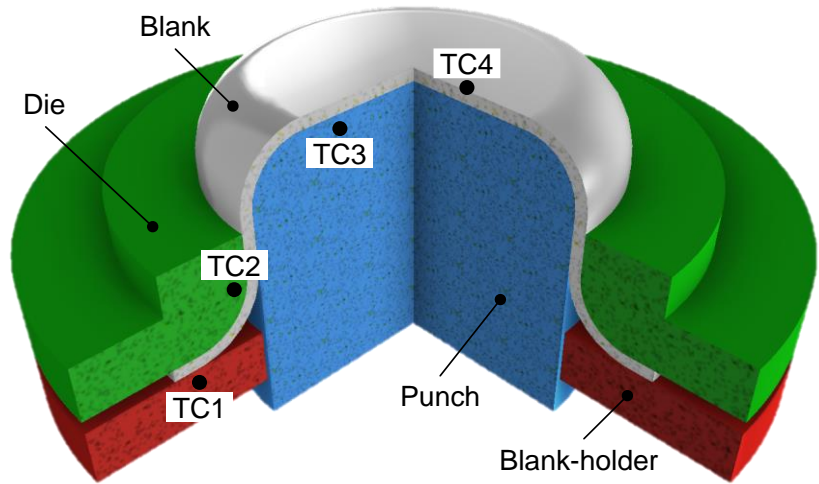

Figure 1: Scheme of the warm deep drawing of a cylindrical cup and location of the four thermocouples (TC).

Table 1: Main dimensions of the forming tools used in the warm deep drawing of a cylindrical cup [mm].

| Die | | | Punch | | Blank-holder |
|---|---|---|---|---|---|
| Opening diameter | Corner radius | Height | Diameter | Corner radius | Opening diameter |
| 35.25 | 5.0 | 8.75 | 33.0 | 5.0 | 33.6 |

Several experimental and numerical studies indicate that the limiting drawing ratio (LDR) of the aluminium alloys can be significantly improved by warm forming (Kurukuri et al., 2009; Li and Ghosh, 2004; Takuda et al., 2002), particularly when temperature gradients are created. For example, the LDR increased from 2.1 to 2.6 by heating the flange and cooling the punch, considering the deep drawing of an AA5754-O aluminium cylindrical cup (van den Boogaard et al., 2001). Therefore, in this benchmark, both the die and the blank-holder are heated up to the test temperature via inserted electrical heating rods (resistance coil and copper plate) (Laurent et al., 2015) while the punch is cooled by water flow (axial channels) (Manach et al., 2016), generating a thermal gradient along the cup wall. The blank is heated between the die and the blank-holder until a uniform temperature is achieved, before forming.

The material of the blank was taken from a rolled sheet of AA5086-H111 aluminium alloy with 0.8 mm of thickness, which is commonly used in the automotive industry (Hirsch, 2011; Toros et al., 2008). Since the circular blank presents a diameter of 60 mm, the drawing ratio of the present cylindrical cup forming is about 1.8. In the experimental deep drawing operation, the punch velocity was 5 mm/s and the blank-holder force was set to 5 kN, which is maintained until the cup is fully drawn (Manach et al., 2016). At the beginning of each deep drawing test, a high temperature lubricant (Jelt Oil) is applied on both sides of the blank. The force–displacement curve of the punch and the blank-holder were recorded during the tests.

In order to measure the temperature evolution of each forming tool and blank, four thermocouples (type K) were used, whose positions are schematically shown in Figure 1. The position of each one (TC1-blank-holder; TC2-die; TC3-punch and TC4-blank) is defined in Table 2, using its radial position and distance to the contact surface (Manach et al., 2016). The temperature distribution on the tools is assumed uniform along the circumferential direction.

Table 2: Position of the thermocouples (type K) used to measure the temperature of each forming tool (TC1, TC2 and TC3) and blank (TC4) during the forming operation (Manach et al., 2016).

| Thermocouple | TC1 | TC2 | TC3 | TC4 |
|---|---|---|---|---|
| Radial distance [mm] | 23.65 | 19.2 | 8.0 | 5.75 |
| Distance to the surface [mm] | 1.0 | 1.0 | 1.5 | Die side |

In addition to the deep drawing operation at room temperature (25ºC), two different non-isothermal heating conditions were considered in the warm forming operation, i.e. both the die and the blank-holder were heated (150ºC and 240ºC), while the punch was cooled (Manach et al., 2016). Five forming tests under identical process conditions were performed in order to check the reproducibility. The experimental temperature evolution of the forming tools (die, blank-holder and punch) and blank is presented in Figure 2, for the two non-isothermal warm deep drawing conditions. In both cases, the temperature of the die and the blank-holder is approximately equal and constant during the forming process, while the punch temperature presents an increase due to the contact with the warm blank. On the other hand, the blank temperature (measured in the bottom of the cup by the thermocouple TC4) presents a significant decrease at the beginning due to the contact with the cold punch, as highlighted in Figure 2. Nevertheless, the temperature difference between the punch and the die/blank-holder generates a temperature gradient from the bottom to the flange of the cylindrical cup (Laurent et al., 2015).

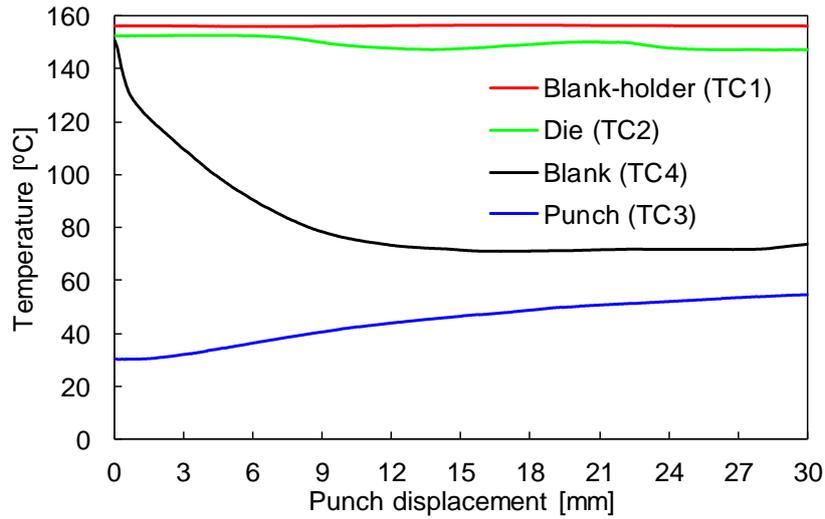

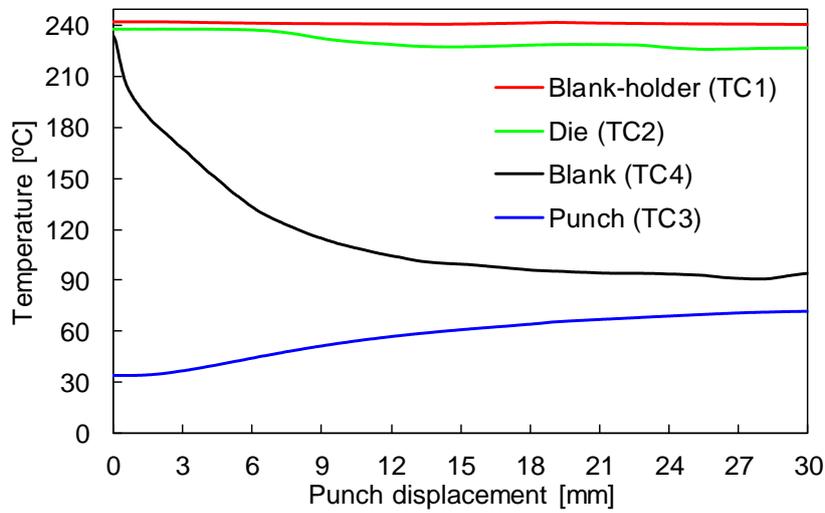

Figure 2: Experimental temperature evolution of the forming tools (blank-holder, die and punch) and blank, considering different values for the initial temperature of the blank: (a) 150ºC; (b) 240ºC (Manach et al., 2016). Temperatures measured by thermocouples placed according to Figure 1 and Table 2.

## 2.2. Split-ring test

The split-ring test, originally proposed by (Demeri et al., 2000) was adopted by (Manach et al., 2016) to predict the effect of temperature on the springback. The experimental residual stress state is evaluated by measuring the opening of a ring cut from the sidewall of the cylindrical cup (Laurent et al., 2010, 2009), as illustrated in Figure 3. Since the cup temperature at the end of the warm forming operation is at least 75ºC (see Figure 2), the cups are left to cool naturally in the air during several hours before the cutting operation. The rings (5 mm height) were cut perpendicularly to the revolution axis at 7 mm from the bottom of the cup, as shown in Figure 3 (a). Both the cutting and splitting (carried out along the axial direction in the RD) were performed by a wire electro-erosion machine (Manach et al., 2016). The measurements of the ring opening were performed along straight lines connecting the two ends of the split ring (see Figure 3 (b)), providing an indirect measure of springback.

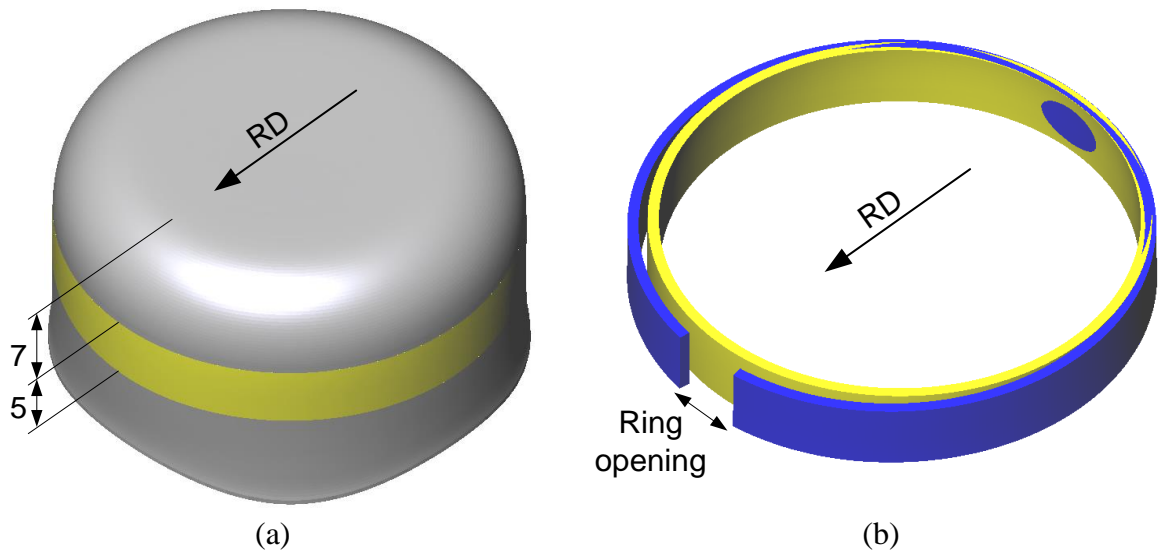

(a)            (b)

Figure 3: Scheme of the split-ring test carried out on the cylindrical cup: (a) location and dimensions (mm) of the ring; (b) opening measured after split.

## 2.3. Thermo-mechanical characterization of the AA5086

In order to identify the material parameters of the constitutive models used in the finite element model, the material (AA5086-H111) was characterized by the benchmark committee (Manach et al., 2016) under different conditions and strain paths. Several uniaxial tensile tests were performed under isothermal conditions at 25ºC (room temperature), 150ºC and 240ºC. The tensile tests at warm temperature were carried out on a Gleeble 3500 testing machine, where the specimen is heated by Joule effect (Coër et al., 2011). The experimental uniaxial tensile stress–strain curves at different temperatures for the RD are presented in Figure 4, for three distinct values of crosshead velocity. This material exhibit serrated flow behaviour at room temperature caused by the Portevin-Le Chatelier effect. However, this effect vanishes for temperatures above 200ºC (Bernard et al., 2016). Increasing the test temperature leads to a decrease of the flow stress and increases the elongation at failure, as highlighted in Figure 4.

Nevertheless, it should be mentioned that the stress–strain curves presented in this figure are represented assuming a uniaxial stress state beyond the onset of necking.

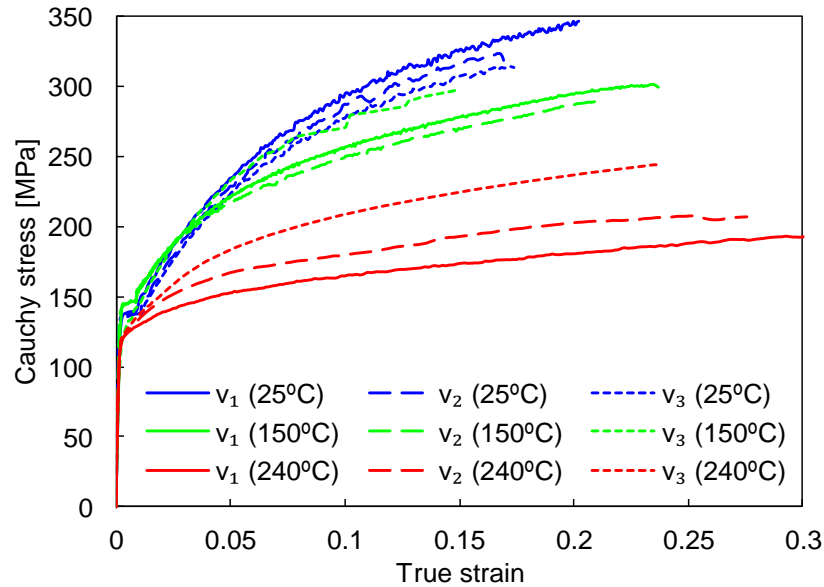

Figure 4: Experimental uniaxial tensile stress–strain curves as a function of the crosshead velocity (three values) and the temperature (three values) for the RD (Manach et al., 2016).

The control of the crosshead velocity is difficult in the Gleeble 3500 testing machine. Thus, the value of the strain rate is not constant during the entire test, as shown in Figure 5 for the three temperatures analysed. In fact, for warm temperatures it is possible to observe an increase of the strain rate, which can be related with the thermal gradient present in tensile specimen (Coër et al., 2011). Nevertheless, it is possible to identify three average distinct values for the strain rate, denoted by $v_1 \approx 0.001$ s$^{-1}$, $v_2 \approx 0.01$ s$^{-1}$ and $v_3 \approx 0.1$ s$^{-1}$. Regarding the strain rate effect in this aluminium alloy, its sensitivity is more pronounced at warm temperatures than at room temperature (see Figure 4), which was also observed by (Abedrabbo et al., 2007) for two aluminium alloys: AA5182-O and AA5754-O. Besides, at room temperature, the flow stress decreases with the increase of the strain rate (negative strain rate sensitivity) due to the Portevin-Le Chatelier effect, while the positive strain rate sensitivity dominates the plastic deformation at 240ºC. For the temperature of 150ºC, the material presents a negative strain rate sensitivity, between $v_1$ and $v_2$, which becomes positive, between $v_2$ and $v_3$. This complex strain rate sensitivity behaviour was modelled by (Kabirian et al., 2014), capturing the transition from negative to positive strain rate sensitivity with the temperature rise.

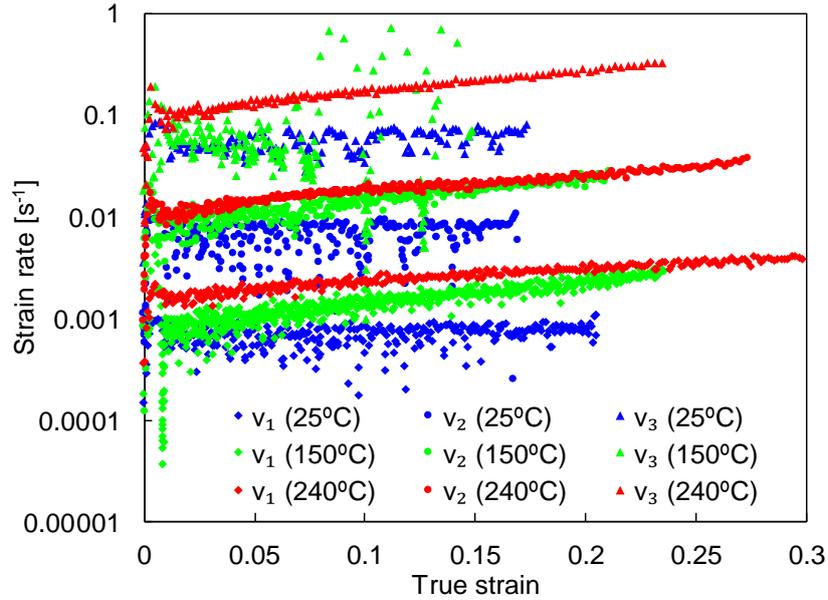

Figure 5: Experimental strain rate of the uniaxial tensile tests performed at different temperatures for the RD (Manach et al., 2016).

The influence of the plastic anisotropy on the experimental stress–strain curves is presented in Figure 6 for uniaxial tensile tests carried out at the average strain rate $v_1$ in three different directions with the RD (Manach et al., 2016). For all isothermal temperature conditions, the flow stress is higher in the RD when compared with the two other directions. In fact, for this aluminium alloy the flow stress measured at DD and TD is identical, particularly at room temperature and at 240ºC (see Figure 6). For the three directions examined, the stress–strain curves comprise a yield plateau both at room temperature and at 150ºC, which is consequence of the Piobert–Lüders phenomenon (de Codes et al., 2011; Mazière et al., 2016).

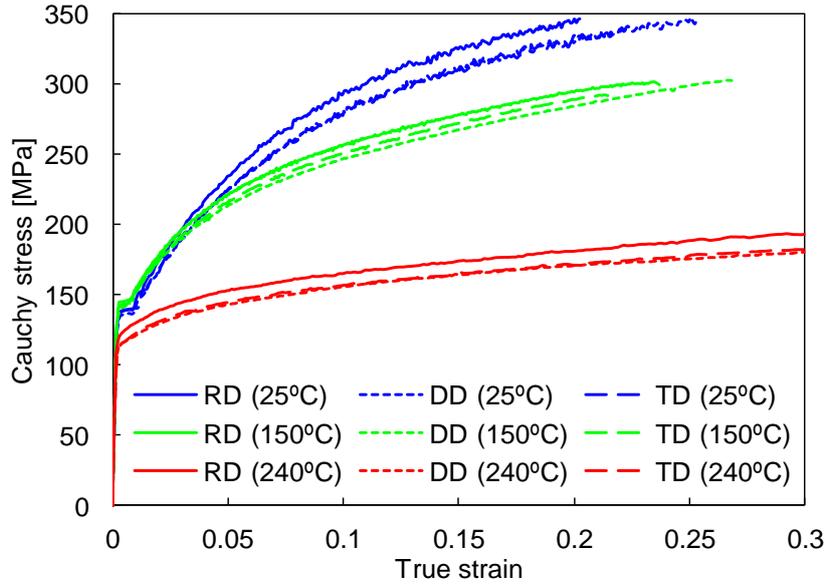

Figure 6: Experimental uniaxial tensile stress–strain curves as a function of the orientation with respect to the RD and the test temperature (tests performed at strain rate $v_1$) (Manach et al., 2016).

The variation of the plastic anisotropy coefficients $r_0$, $r_{45}$ and $r_{90}$ with respect to temperature (Manach et al., 2016) is presented in Table 3. These values were obtained from the uniaxial tensile tests presented in Figure 6, which were carried out using the lower average value of strain rate ($v_1 \approx 0.001$ s$^{-1}$). Globally, the $r$-values decrease with the temperature rise, suggesting that the formability of this aluminium alloy will deteriorate at warm temperatures. Opposite behaviour was observed by (Abedrabbo et al., 2007) for two aluminium alloys: AA5182-O and AA5754-O. In addition to the uniaxial tensile tests, the benchmark committee provided data of monotonous and reverse shear tests at different temperatures (up to 150ºC) and a biaxial test (hydraulic bulge) at room temperature (Manach et al., 2016). However, since this information was not taken into account in this study it is not reported here.

Table 3: Experimental anisotropy coefficients depending on the temperature (Manach et al., 2016).

| Temperature [ºC] | $r_0$ | $r_{45}$ | $r_{90}$ |
|---|---|---|---|
| 25 | 0.71 | 1.08 | 0.73 |
| 150 | 0.63 | 0.97 | 0.66 |
| 240 | 0.60 | 0.88 | 0.67 |

## 3. Thermo-mechanical constitutive model

The thermo-mechanical nature of the warm forming process requires the adoption of numerical models able to describe accurately the mechanical behaviour of the material at different values of temperature. In the present study, the deformation of the metallic sheet is described by a rate-independent thermo-elasto-plastic material model. The mechanical behaviour of the aluminium alloy is assumed non-linear and anisotropic in the plastic domain (orthotropic plasticity). Hence, the plastic response is defined by a hardening law and a yield function, which are connected through an associated no-viscous flow rule.

### 3.1. Hockett–Sherby hardening law

The phenomenological Hockett–Sherby hardening law (Hockett and Sherby, 1975) is adopted to describe the flow stress of the AA5086-H111 aluminium alloy at different temperatures. Since the temperature affects significantly the flow stress (see Figure 4), an appropriate constitutive model is required. Hence, the isotropic work hardening behaviour, which describes the evolution of the flow stress with plastic work, is modelled by (Laurent et al., 2015):

$$Y = Y_0 + Q\{1 - \exp(-b(\bar{\varepsilon}^{\mathrm{p}})^n)\}, \tag{1}$$

where $\bar{\varepsilon}^{\mathrm{p}}$ denotes the equivalent plastic strain, $Y_0$ is the initial value of the yield stress and $b$ defines the growth rate of the yield surface. The temperature dependence is incorporated in the parameters $Q$ and $n$, which are defined as function of the temperature $T$. The evolution of the maximum size of the yield surface is given by:

$$Q = Q_0 + a_1\left\{1 - \exp\left(a_2 \frac{T}{T_{\mathrm{m}}}\right)\right\}, \tag{2}$$

where $Q_0$, $a_1$ and $a_2$ are material parameters. The strain hardening index $n$, which evolves linearly with the temperature, is given by:

$$n = n_0 - n_1\left(\frac{T}{T_{\mathrm{m}}}\right), \tag{3}$$

where $n_0$ and $n_1$ are material parameters, while $T_{\mathrm{m}}$ denotes the melting temperature.

The initial yield stress predicted by this constitutive model is temperature-independent. Nevertheless, considering the aluminium alloy under analysis and ignoring the yield plateau, the variation of the initial yield stress with the temperature can be considered negligible in comparison with the variation of the ultimate tensile strength, as shown in Figure 4. Besides, although this aluminium alloy exhibits an evident strain rate sensitivity (see Figure 4), the hardening law adopted is unable to model this effect. In order to overcome this drawback, the identification procedure is performed considering the experimental stress–strain curves obtained at an average strain rate close to the value arisen in the deep drawing process. Although the strain rate value assessed during the forming operation is not constant nor uniform (see Section 4.3), most of the plastic strain occurs for a strain rate value close to $v_3 \approx 0.1$ s$^{-1}$. Thus, only the experimental uniaxial tensile tests carried out for the RD, at the higher average strain rate and different temperatures (25ºC, 150ºC and 240ºC) are considered in the identification procedure, as shown in Figure 7.

The identification procedure adopted is based on the minimization of an error function, which evaluates the difference between the numerical and the experimental stress values (Dasappa et al., 2012). Therefore, the optimization problem consists in identifying the set of material parameters $\vartheta$ that minimizes the following error function:

$$F(\vartheta) = \sum_{i=1}^{3} \left\{ \sum_{k=1}^{n_i} \left( \sigma_{\exp}^i(k) - \sigma_{\text{num}}^i(k) \right)^2 \right\}, \qquad (4)$$

where the first sum represents the three uniaxial tensile tests carried out at different isothermal conditions, $n_i$ is the number of measured points at each temperature and $\sigma$ denotes the tensile stress value. The subscripts "exp" and "num" denote the experimental and numerical data, respectively. Since the stress values are non-normalized in Eq. (4), the importance of high stress values is enhanced, which leads to a better fit of the experimental stress–strain curves for large values of plastic strain (closer to the ones observed in the cylindrical cup). The obtained material parameters for the isotropic hardening law are listed in Table 4. The value of the melting temperature $T_m$ was assumed as fixed in the optimization procedure.

Table 4: Material parameters used in the Hockett–Sherby hardening law to describe the AA5086 aluminium alloy.

| $Y_0$ [MPa] | $Q_0$ [MPa] | $B$ | $a_1$ [MPa] | $a_2$ | $n_0$ | $n_1$ | $T_m$ [ºC] |
|---|---|---|---|---|---|---|---|
| 107.07 | 262.49 | 6.989 | 2.834 | 9.459 | 0.819 | 0.293 | 600 |

The comparison between experimental and numerical stress–strain curves is presented in Figure 7. The thermo-mechanical behaviour of this aluminium alloy is accurately described by this hardening law, since the numerical results are in good agreement with the experimental ones. For the higher average strain rate value, experimentally measured in the tensile tests ($v_3 \approx 0.1$ s$^{-1}$), the flow stress measured at 25ºC and 150ºC is similar. Thus, the mechanical behaviour of this material is almost thermal insensitive in this temperature range.

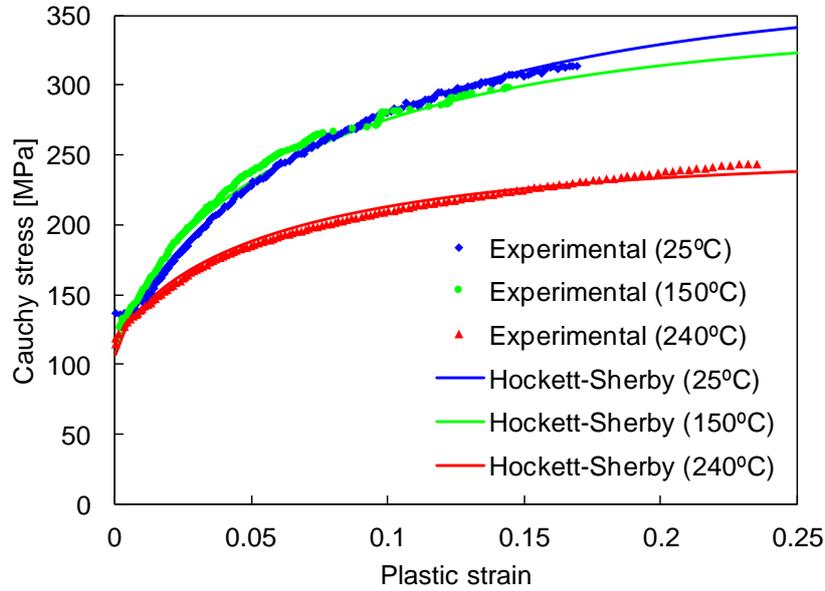

Figure 7: Experimental and numerical uniaxial tensile stress–strain curves (performed at strain rate $v_3$ in the RD) for three different temperatures.

## 3.2. Yield function

In order to model the anisotropic behaviour of this aluminium alloy, the classical Hill'48 yield criterion (Hill, 1948) is compared with the non-quadratic yield criterion proposed by (Barlat et al., 1991), which is particularly adequate for aluminium alloys. Although the Hill'48 yield criterion is successfully used in the sheet metal forming simulation of steels (Kim et al., 2013), it is known to be inadequate for describing the behaviour of aluminium alloys (Tardif, 2012).

The yield criteria adopted are considered temperature-independent, despite the variation of the plastic anisotropy parameters with the temperature (see Table 3). According to the Hill'48 yield criterion, defined in the appropriate orthogonal rotating orthotropic frame, the equivalent stress $\bar{\sigma}$ is expressed by:

$$\bar{\sigma}^2 = F(\sigma_{22} - \sigma_{33})^2 + G(\sigma_{33} - \sigma_{11})^2 + H(\sigma_{11} - \sigma_{22})^2 + \\ + 2L(\sigma_{23})^2 + 2M(\sigma_{13})^2 + 2N(\sigma_{12})^2, \tag{5}$$

where $F$, $G$, $H$, $L$, $M$ and $N$ are the parameters that describe the anisotropy of the material, while $\sigma_{11}$, $\sigma_{22}$, $\sigma_{33}$, $\sigma_{23}$, $\sigma_{13}$ and $\sigma_{12}$ are the components of the Cauchy stress tensor defined in the orthotropic frame.

The yield criterion proposed by (Barlat et al., 1991) is an extension to orthotropy of the Hosford isotropic yield criterion (Hosford, 1972). This anisotropic yield criterion is defined by the following non-quadratic yield function:

$$|S_1 - S_2|^m + |S_2 - S_3|^m + |S_3 - S_1|^m = 2\bar{\sigma}^m, \tag{6}$$

where $S_1$, $S_2$ and $S_3$ are the eigenvalues of the isotropic stress tensor obtained from the linear transformation applied to the Cauchy stress tensor $\mathbf{S} = \mathbf{L} : \boldsymbol{\sigma}$. The exponent $m$ is connected to the material crystallographic structure, providing the shape of the yield surface, i.e. $m=6$ for BCC and $m=8$ for FCC metals (Logan and Hosford, 1980). The six anisotropy coefficients are contained within the linear transformation matrix $\mathbf{L}$, given by:

$$\mathbf{L} = \frac{1}{3}\begin{bmatrix} (c_2 + c_3) & -c_3 & -c_2 & 0 & 0 & 0 \\ -c_3 & (c_3 + c_1) & -c_1 & 0 & 0 & 0 \\ -c_2 & -c_1 & (c_1 + c_2) & 0 & 0 & 0 \\ 0 & 0 & 0 & 3c_4 & 0 & 0 \\ 0 & 0 & 0 & 0 & 3c_5 & 0 \\ 0 & 0 & 0 & 0 & 0 & 3c_6 \end{bmatrix}, \tag{7}$$

where $c_1$, $c_2$, $c_3$, $c_4$, $c_5$ and $c_6$ are the anisotropy parameters.

Typically, the parameters of the Hill'48 yield criterion are evaluated based on the anisotropy coefficients (Dasappa et al., 2012). Accordingly, the anisotropy parameters are assessed through the following relations:

$$\frac{H}{G} = r_0; \quad \frac{F}{G} = \frac{r_0}{r_{90}}; \quad \frac{N}{G} = \left(r_{45} + \frac{1}{2}\right)\left(\frac{r_0}{r_{90}} + 1\right), \tag{8}$$

where $r_0$, $r_{45}$ and $r_{90}$ are the $r$-values obtained experimentally by uniaxial tensile tests carried in the RD, DD and TD, respectively. Since the material parameters for the hardening law (see Table 4) were identified using the stress–strain curves in the RD, the condition $G+H=1$ is also imposed. In order to capture the temperature effect on the material anisotropic behaviour, the set of parameters is evaluated for each temperature (25ºC, 150ºC and 240ºC) according to the $r$-values presented in Table 3. The anisotropy parameters identified using this approach are labelled Hill'48-R and are presented in Table 5, for each temperature evaluated. The sheet is assumed isotropic through the thickness, leading to $L=M=1.5$.

Table 5: Material parameters of the Hill'48-R yield criterion for different isothermal temperature conditions.

| Temperature [ºC] | $F$ | $G$ | $H$ | $L$ | $M$ | $N$ |
|---|---|---|---|---|---|---|
| 25 | 0.5688 | 0.5848 | 0.4152 | 1.500 | 1.500 | 1.8226 |
| 150 | 0.5856 | 0.6135 | 0.3865 | 1.500 | 1.500 | 1.7627 |
| 240 | 0.5597 | 0.6250 | 0.3750 | 1.500 | 1.500 | 1.6349 |

Since the stress–strain curves present a yield plateau at room temperature and at 150ºC (see Figure 6), the definition of the initial yield stress can be dubious. Therefore, the stress–strain curves of the uniaxial tensile tests were normalized by the flow stress of the material in the RD. Figure 8 presents the evolution of the normalized uniaxial stress values, evaluated at different temperatures and considering the lower average strain rate value. The material anisotropic behaviour is evaluated at the lowest strain rate due to lack of experimental data, namely tensile tests at the highest strain rate for different directions with the rolling direction. Nevertheless, previous results for a 5xxx aluminium alloy show that the anisotropy coefficients are relatively insensitive to the strain rate (Rahmaan et al., 2016). For all temperature values studied, this aluminium alloy exhibits a flow stress curve higher in the RD and lower in the DD, i.e. the values of normalized uniaxial stress are less than one. Besides, the stress ratios are roughly constant during the test, as shown in Figure 8, indicating that the stress ratios are insensitive to the true strain, particularly at 150ºC.

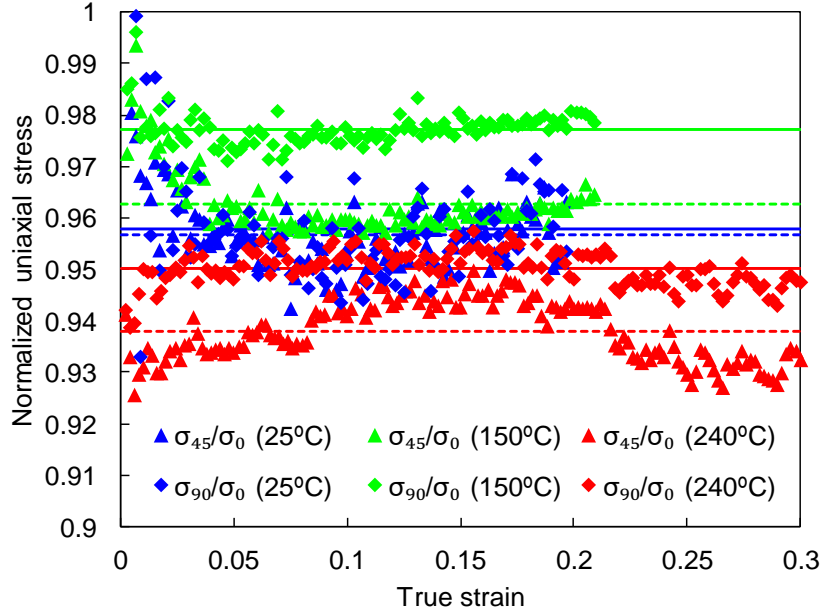

Figure 8: Normalized uniaxial stress values experimentally evaluated in the uniaxial tensile tests performed at different temperatures and considering the strain rate $v_1$.

In the present study, the average value of the normalized uniaxial flow stress values $\sigma_0/\sigma_0 = 1$, $\sigma_{45}/\sigma_0$ and $\sigma_{90}/\sigma_0$ is used in the identification procedure to assess the anisotropy parameters. Hence, the initial yield stress given by the hardening law (107.07 MPa) is used to restore the absolute value of the initial uniaxial yield stress in the three directions. For the Hill'48 yield criterion, considering $\sigma_0$ as the reference value for the initial yield stress (hardening law identified using the stress–strain curves in the RD), the in-plane anisotropy parameters can also be assessed by:

$$G + H = 1; \quad \sqrt{F+H} = \frac{\sigma_0}{\sigma_{90}}; \quad \sqrt{F+G+2N} = \frac{2\sigma_0}{\sigma_{45}}, \tag{9}$$

where $\sigma_0$, $\sigma_{45}$ and $\sigma_{90}$ are the yield stresses obtained experimentally by uniaxial tensile tests carried in the RD, DD and TD, respectively. In order to obtain four equations (four in-plane anisotropic coefficients), the *r*-value obtained experimentally by the uniaxial tensile test carried in the RD is also used, i.e. the first relation presented in Eq. (8). The set of material parameters is evaluated for each temperature (25ºC, 150ºC and 240ºC) in order to describe the influence of the temperature on the material anisotropic behaviour. The anisotropy parameters identified using this approach are labelled Hill'48-S and are presented in Table 6.

Table 6: Material parameters of the Hill'48-S yield criterion for different isothermal temperature conditions.

| Temperature [°C] | $F$ | $G$ | $H$ | $L$ | $M$ | $N$ |
|---|---|---|---|---|---|---|
| 25 | 0.6750 | 0.5848 | 0.4152 | 1.500 | 1.500 | 1.5547 |
| 150 | 0.6608 | 0.6134 | 0.3866 | 1.500 | 1.500 | 1.5216 |
| 240 | 0.7303 | 0.6250 | 0.3750 | 1.500 | 1.500 | 1.5848 |

The material parameters involved in the yield criterion proposed by (Barlat et al., 1991) are identified taking into account both the experimental yield stresses and the Lankford coefficients (listed in Table 3). Additionally, the biaxial yield stress at room temperature (108 MPa) is also used in the identification procedure. This value is assumed temperature-independent in the material parameters identification, due to the lack of experimental data at warm temperatures. The identification procedure adopted is based on the minimization of an error function that evaluates the difference between the predicted and the experimental values, which was implemented in DD3MAT in-house code (Barros et al., 2016). Such as in Hill'48 yield criterion, the set of parameters for the Barlat'91 yield criterion was evaluated for each temperature (25°C, 150°C and 240°C), and are presented in Table 7. The parameters defining the anisotropic behaviour through-thickness direction are assumed isotropic, i.e. $c_4=c_5=1.0$ (Tardif, 2012).

Table 7: Material parameters of the Barlat'91 yield criterion ($m=8$) for different isothermal temperature conditions.

| Temperature [°C] | $c_1$ | $c_2$ | $c_3$ | $c_4$ | $c_5$ | $c_6$ |
|---|---|---|---|---|---|---|
| 25 | 1.0392 | 1.0454 | 0.9478 | 1.0000 | 1.0000 | 1.0583 |
| 150 | 1.0402 | 1.0536 | 0.9228 | 1.0000 | 1.0000 | 1.0405 |
| 240 | 1.0502 | 1.0830 | 0.9355 | 1.0000 | 1.0000 | 1.0406 |

The in-plane distribution of the *r*-values and the initial yield stresses is presented in Figure 9, comparing the distributions predicted by the yield criteria (Hill'48-R, Hill'48-S and Barlat'91) with the experimental results. None of the yield functions considered in the present study is able to fit simultaneously the yield stress and the *r*-value in-plane distributions. The adoption of more advanced yield functions, such as (Barlat et al., 2005) or a non-associated flow rule (Park and Chung, 2012), allows to overcome this drawback. However, advanced yield functions require a larger set of experimental data to identify the anisotropy parameters, while the adoption of non-associated flow rules is prone to the computational complexity of stress integration procedure (Hippke et al., 2017). Both the Hill'48-R and the Barlat'91 yield criteria provide identical distributions for the *r*-value, fitting accurately the experimental ones (see Figure 9 (a)). Nevertheless, when the anisotropy parameters of the Hill'48 yield criterion are evaluated using three yield stresses and one *r*-value (Hill'48-S), the experimental *r*-values are globally underestimated, particularly the plastic anisotropy coefficient $r_{45}$. On the other hand, this set of anisotropy parameters (Hill'48-S) leads to a perfect fit of the experimental yield stresses. Globally, the Barlat'91 yield criterion leads to a good description of both yield stress and *r*-value in-plane distributions, as shown in Figure 9. Regarding the influence of the temperature on the material anisotropic behaviour, the in-plane distribution of the initial yield stress is roughly temperature insensitive, particularly for the Barlat'91 yield criterion. Nevertheless, the *r*-value decreases substantially with the temperature rise, particularly when it is modelled by the Hill'48-R and the Barlat'91 yield criteria, as shown in Figure 9 (a).

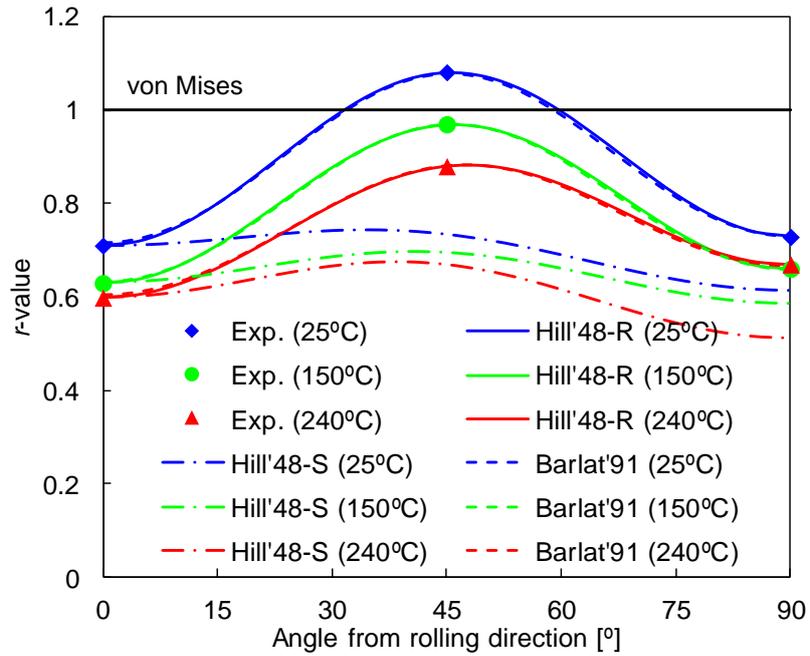

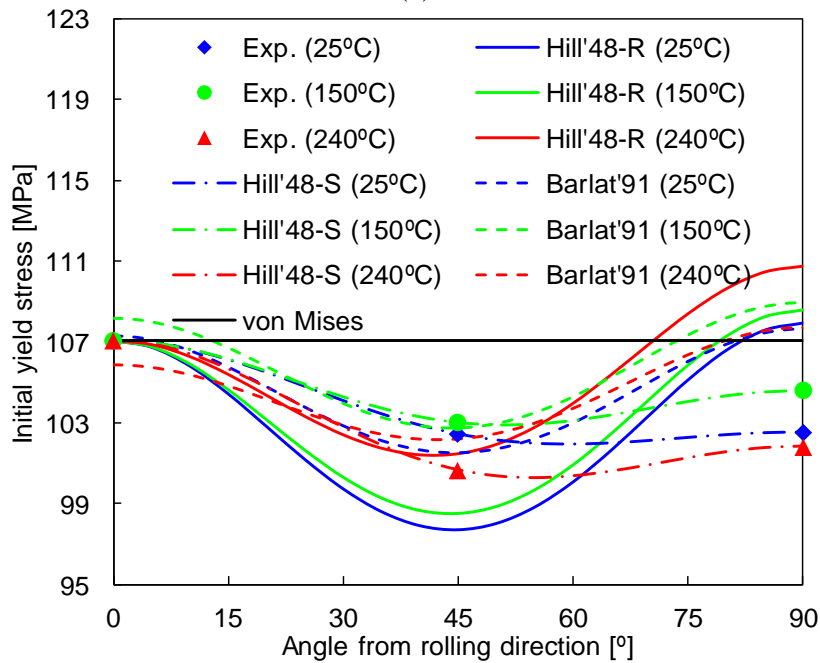

Figure 9: Comparison between experimental and predicted anisotropy: (a) *r*-value; (b) initial uniaxial yield stress.

The yield surface predicted by each yield criteria is displayed in Figure 10 for the $\sigma_{11} - \sigma_{22}$ plane. The influence of the temperature conditions on the yield loci is nearly imperceptible, except when considering the Hill'48-S evaluated at 240ºC. Nevertheless, the anisotropic yield criteria provide distinct shapes for the yield surface, being both inside the von

Mises yield locus for the biaxial stress path. The balanced biaxial stress predicted by the Hill'48 yield criterion is lower than the one predicted by Barlat'91 yield criterion, particularly when the anisotropy parameters of the Hill'48 yield criterion are evaluated using three yield stresses and one *r*-value (Hill'48-S), as shown in Figure 10.

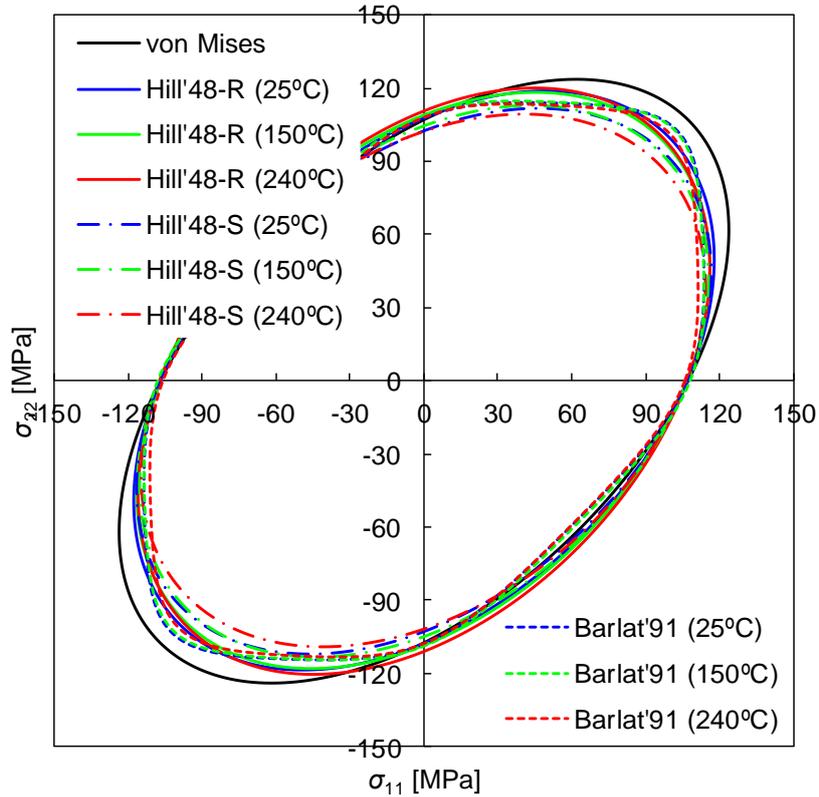

Figure 10: Comparison of different yield surfaces (von Mises, Hill'48 and Barlat'91) evaluated in the $\sigma_{11} - \sigma_{22}$ plane for three different temperatures.

## 4. Finite element model

In order to model the complete warm deep drawing process (including the split-ring test), the simulation is divided into six different stages: (i) heating of the blank within the tools; (ii) deep drawing operation; (iii) cooling of the cylindrical cup; (iv) unloading the cup; (v) cutting the ring and (vi) split ring. The numerical simulations were carried out with the in-house static implicit finite element code DD3IMP (Menezes and Teodosiu, 2000), specifically developed to simulate sheet metal forming processes (Menezes et al., 2011; Oliveira et al., 2008). All simulations were performed on a computer machine equipped with an Intel® Core™ i7-5930K Quad-Core processor (3.5 GHz) and the Windows® 10 (64-bit platform) operating system.

In warm forming processes, the temperature of the blank (distribution and evolution) depends on the temperature of the tools, as well as the heat transfer mechanisms between the tools and the blank. Therefore, in the present study, the thermo-mechanical problem is solved using the staggered coupled strategy proposed by (Martins et al., 2017), currently implemented in the in-house finite element code DD3IMP. Within each time step, the proposed algorithm is divided into two phases, a prediction phase and a correction phase. In both phases the thermal and the mechanical problem are solved, but using different methods and input data. The prediction phase begins with the resolution of the thermal problem, which is solved for a user-defined time-step, adopting the Crank-Nicolson method. In this first phase, the solution of the displacement field is unknown and, consequently, the heat generated by plastic deformation or frictional contact, which plays an important role in the temperature evolution, is also unknown. Therefore, the amount of heat generated is estimated based on the values obtained in the previous time-step (J M P Martins et al., 2016). This allows determining a trial temperature field, which is used to solve the mechanical problem, adopting an explicit method. At the end of the prediction phase, a trial displacement field and a trial temperature field are available, which do not satisfy the equilibrium equations. Therefore, these trial solutions are corrected using an implicit method for the time integration of the respective equation of equilibrium. The mechanical problem is solved first during the correction phase, using the trial temperature field and the trial displacement field. Finally, the thermal problem is solved using the Euler's backward method implicit approach. Table 8 presents the thermal properties adopted in the finite element model, extracted from (Manach et al., 2016). The proposed model takes into account the heat generated by plastic deformation, which corresponds to the fraction of plastic power (90% in the present study) converted into heat s(J M P Martins et al., 2016).

Several experimental studies indicate that the Young's modulus decreases with the temperature rise. The study carried out by (Laurent et al., 2015) on the AA5754-O aluminium alloy shows that the Young modulus decreases from about 66 GPa (room temperature) down to 56 GPa at 300ºC. The influence of the temperature on the Young modulus of the AA5052-H32 aluminium alloy was experimentally assessed by Masubuchi (Masubuchi, 1980), presenting a decrease of around 15 GPa with the increase of the temperature up to 240ºC. Nevertheless, in the present study the elastic properties of the aluminium alloy are considered temperature-independent, allowing to simplify the numerical model. On the other hand, the Young modulus reduces with the accumulated plastic strain (Cleveland and Ghosh, 2002). The experimental measurements performed by (Alghtani, 2015) on a high strength aluminium alloy highlight the Young's modulus degradation, which decreases about 10 GPa for 10% of accumulated plastic strain. Nevertheless, due to lack of experimental data, the elastic behaviour is assumed isotropic and constant in the present study, which is described by the Hooke's law using the parameters listed in Table 8.

Table 8: Mechanical (elastic) and thermal properties of the AA5086 aluminium alloy.

| Young modulus [GPa] | Poisson's ratio | Specific density [kg/m$^3$] | Specific heat [J/kg·°C] | Thermal conductivity [W/m·°C] |
|---|---|---|---|---|
| 71.7 | 0.31 | 2700 | 900 | 220 |

## 4.1. Discretization of blank and tools

Due to geometric and material symmetry conditions, only half model is simulated. This allows to simplify the analysis of the cutting and splitting stages, by just removing the symmetry condition at one end of the ring. The blank is discretized with linear hexahedral finite elements, as shown in Figure 11, using 3 layers of elements through the thickness. The central zone of the blank (flat area of the punch) is discretized by a relatively coarse unstructured mesh, while the remaining zone is discretized with a fine structured mesh. The same finite element mesh is used in the mechanical and thermal problem (Adam and Ponthot, 2005), avoiding the application of mapping methods to exchange information between different discretizations. Nevertheless, full integration is adopted in the thermal problem, while the mechanical problem resorts to the selective reduced integration technique (Hughes, 1980) to avoid volumetric locking.

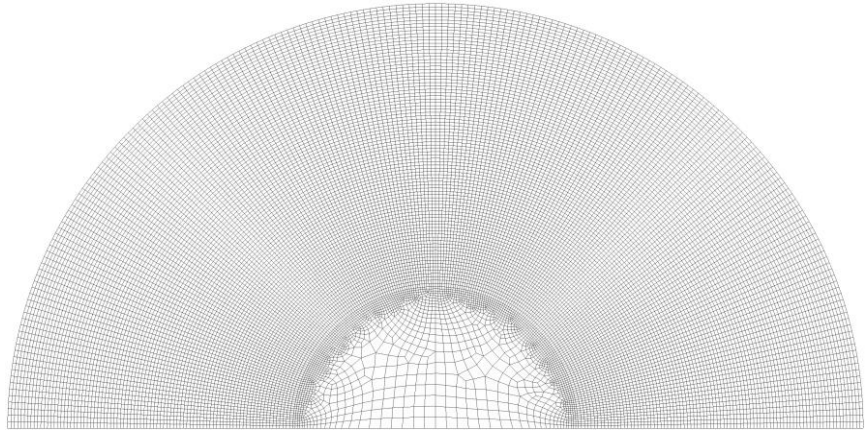

Figure 11: Discretization of half blank using 58,806 solid finite elements (3 layers through the thickness).

The forming tools are usually considered as rigid in the numerical simulation. Thus, in the present study only its surfaces are discretized with Nagata patches (Neto et al., 2014b), as shown in Figure 12. Since the Nagata patch interpolation (Nagata, 2005) recovers the curvature of the surfaces with good accuracy, the numerical model only comprises 369 Nagata patches. The nodal normal vectors required for the smoothing method are evaluated from the IGES file, using the algorithm proposed by (Neto et al., 2013).

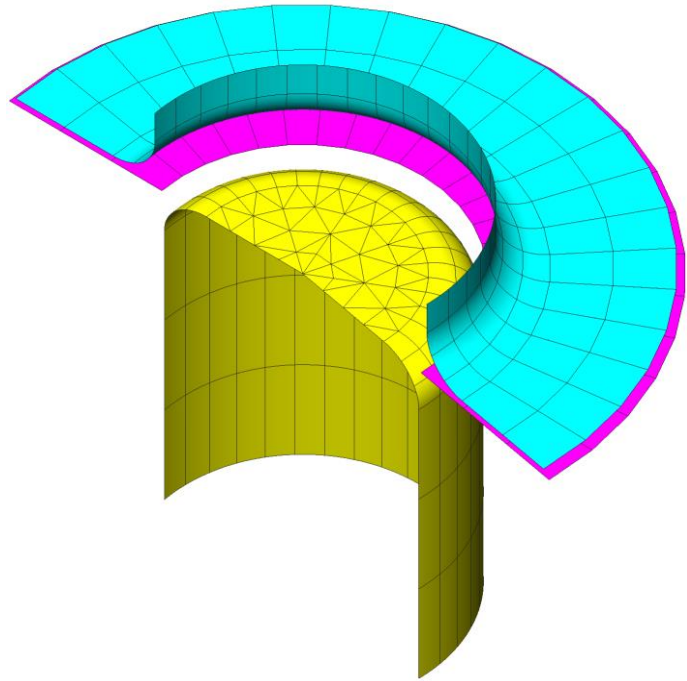

Figure 12: Discretization of the forming tools using 369 Nagata patches.

## 4.2. Thermo-mechanical contact conditions

The friction between the blank and the forming tools is modelled through the classical isotropic Coulomb's law. The value of the friction coefficient used in the numerical simulations, which is considered temperature-independent, was suggested by the benchmark committee (Manach et al., 2016). This value ($\mu = 0.09$) was estimated by (Laurent et al., 2015) through the comparison of numerical results with experimental data, namely the punch force, the thickness and the ear profiles. The frictional contact between the forming tools and the blank generates energy, which is predominantly converted into heat (100% in the present study), producing a temperature increase at the interface (Chen and Chen, 2013). Besides, the fraction of generated energy converted into heat is assumed equally partitioned between the two contacting interfaces. Nevertheless, only the temperature of the blank is affected by frictional heat generation since the temperature of each forming tool is assumed constant in the model (J M P Martins et al., 2016).

Since the warm deep drawing process was carried under non-isothermal conditions (see Section 2.1), the heat transfer between the forming tools and the blank must be taken into account in the numerical model, dictating the cooling of the blank during the forming operation. In the present study, the temperature of the forming tools is assumed uniform and constant during the whole deep drawing operation. This assumption is in agreement with the experimental measurements (see Figure 2), mainly for the die and blank-holder. Therefore, the temperature of all tools in the forming operation at room temperature is 25ºC. On the other hand, during the warm forming operation, both the die and the blank-holder are set at 150ºC and 240ºC. Since the temperature of the punch increases throughout the warm forming operation (see Figure 2), the final value experimentally measured is adopted in the numerical model, i.e. the punch is set at 55ºC and 70ºC. It should be noted that the accuracy of the finite element simulation can be improved by imposing the temperature history presented in Figure 2 on the tools or simply imposing a linear increase evolution, especially for the punch.

In addition to the temperature of the forming tools, the temperature gradient in the blank is controlled by the heat flow across the contacting interfaces, which is commonly defined by the interfacial heat transfer coefficient. Nevertheless, this coefficient is very difficult to evaluate experimentally due to its variation with the contact pressure, the surface temperature difference, among other parameters (Caron et al., 2014; Zhao et al., 2015). Therefore, in the present study, the experimental evolution of the blank temperature (measured in the bottom of the cup) is used to estimate the value of the interfacial heat transfer coefficient. In order to improve the accuracy of the numerical model, the interfacial heat transfer coefficient takes into account the gap distance between bodies, providing a smooth transition between contact status (J M P Martins et al., 2016). Hence, the interfacial heat transfer coefficient is expressed by:

$$h = h_{max} \exp(-mg_n), \tag{10}$$

where $h_{max}$ is the maximum value of the interfacial heat transfer coefficient, $g_n$ denotes the gap distance between the contacting bodies and $m$ is a fitting parameter controlling the rate of the decrease. The heat exchange to the environment (natural convection) is neglected in the proposed model because the forming device is completely closed (Laurent et al., 2015) and the deep drawing operation is performed in less than 6 seconds.

The two parameters involved in the definition of the interfacial heat transfer coefficient, described in Eq. (10), are identified using two distinct strategies. The maximum value of the interfacial heat transfer coefficient was selected from values reported in the literature (Caron et al., 2013, 2014). Afterwards, the fitting parameter $m$ was optimized using numerical results of the warm deep drawing process (150ºC and 240ºC), minimizing the difference to the experimental data. The comparison between experimental and numerical temperature evolution, measured in the bottom of the cup (TC4), is presented in Figure 13 for the three different temperature conditions. The numerical simulations were carried out considering the Barlat'91 yield criterion. For all temperature conditions considered throughout the forming operation, the numerical distribution of the blank temperature is in very good agreement with the experimental results, particularly up to 18 mm of punch displacement.

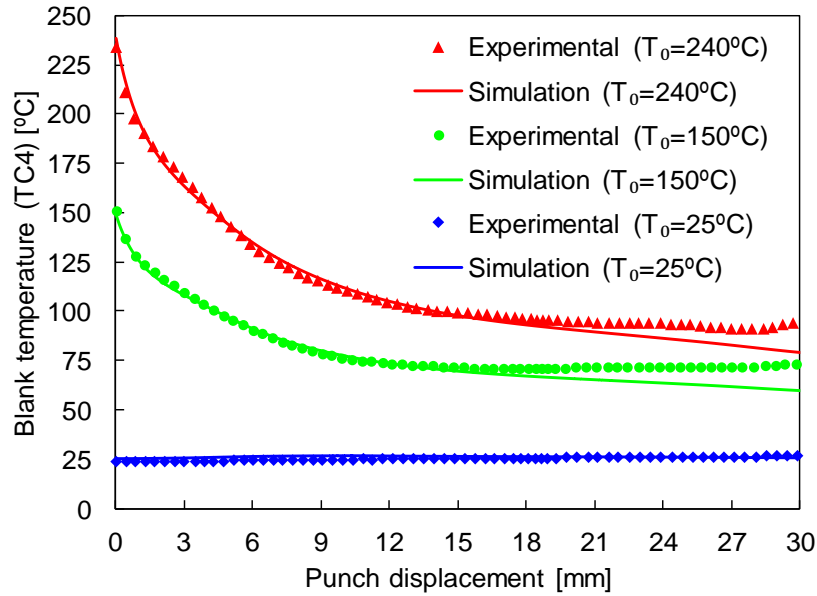

Figure 13: Comparison between experimental and numerical blank temperature (TC4) evolution for three different temperature conditions. Results obtained with the Barlat'91 yield criterion.

The value of each parameter involved in the definition of the interfacial heat transfer coefficient is $h_{max} = 4500$ W/m$^2 \cdot$°C and $m = 8.59$, which was calibrated according to the temperature values recorded during the cup forming experiments. Considering the present model, the value of the interfacial heat transfer coefficient decreases to about 2000 W/m$^2 \cdot$°C when the clearance between the blank and the tool surface is 0.1 mm. In fact, for gap distances higher than 0.5 mm the value of the interfacial heat transfer coefficient is lower than 100 W/m$^2 \cdot$°C (typical value for natural convection coefficient).

### 4.3. Temperature distribution

The temperature distribution in the cylindrical cup cannot be experimentally evaluated during the deep drawing operation due to the visibility constraints imposed by the forming tools. Nevertheless, heating the die/blank-holder and cooling the punch during the warm forming creates a temperature gradient from the bottom to the flange of the cup. The temperature distribution predicted by the numerical simulation (Barlat'91 yield criterion) at the instant corresponding to 15 mm of punch displacement is presented in Figure 14, comparing the two non-isothermal warm deep drawing conditions. Although the material presents anisotropic behaviour, the temperature distribution in the cylindrical cup is approximately axisymmetric, presenting the minimum value in the bottom centre of the cup and the maximum in the flange. In fact, the temperature of the flange is approximately the temperature assigned to the die and the blank-holder due to the large contact area. Since the aluminium presents high thermal conductivity (see Table 8), the temperature gradient through the thickness is negligible (J. M. P. Martins et al., 2016), as shown in Figure 14.

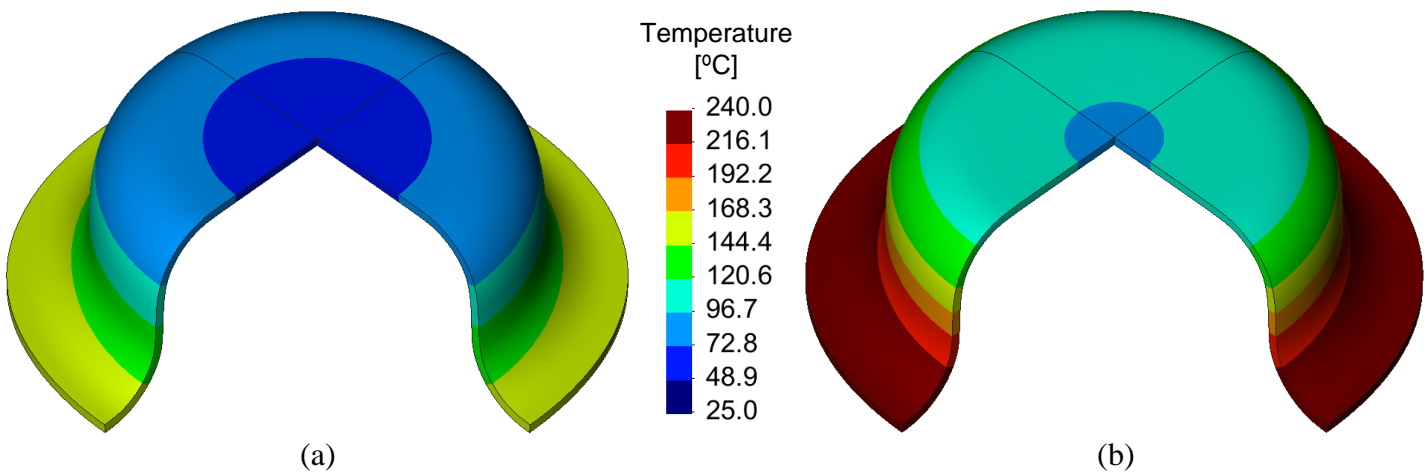

Figure 14: Temperature distribution numerically predicted for the instant corresponding to 15 mm of punch displacement, considering different values for the initial temperature of the blank: (a) 150ºC; (b) 240ºC. Results obtained with the Barlat'91 yield criterion.

In order to analyse the temperature evolution in distinct regions of the cylindrical cup, Figure 15 (a) presents the temperature evolution of a point initially located in the flange (5 mm from the perimeter). Considering the warm deep drawing conditions, the temperature of the point in the flange and die radius areas presents only a slight decrease due to the continuous contact with the heated tools. On the other hand, the slight temperature increase observed in the deep drawing operation at room temperature is induced by the heat generated by plastic deformation and frictional contact. Since the distance between the analysed point and the heated tools (die and blank-holder) is increasing, the temperature decreases when it arrives at the cup wall (converging to the punch temperature), as shown in Figure 15 (a).

Since this aluminium alloy is strain rate sensitive, particularly at warm temperatures (see Figure 4), it is important to assess the strain rate arising in the present deep drawing process (cylindrical cup). The evolution of the plastic strain rate, evaluated in a point initially located in the flange (5 mm from the perimeter), is shown in Figure 15 (b) for three different temperature conditions. Although the temperature has a large impact on the flow stress, its influence on the plastic strain rate is insignificant. The plastic strain rate increases from about 0.01 s$^{-1}$ to 0.1 s$^{-1}$ during the sliding of the point over the die (flange region). Then, this value of plastic strain rate is kept roughly constant during the sliding of the point over the die radius. Finally, the plastic strain rate decreases abruptly when the point reach the cup wall, as shown in Figure 15 (b). This results indicate that the identification of the material parameters for the hardening law (see Table 4) using the uniaxial tensile tests carried out at higher average value of strain rate ($v_3 \approx 0.1$ s$^{-1}$) can provide an accurate description of the flow stress.

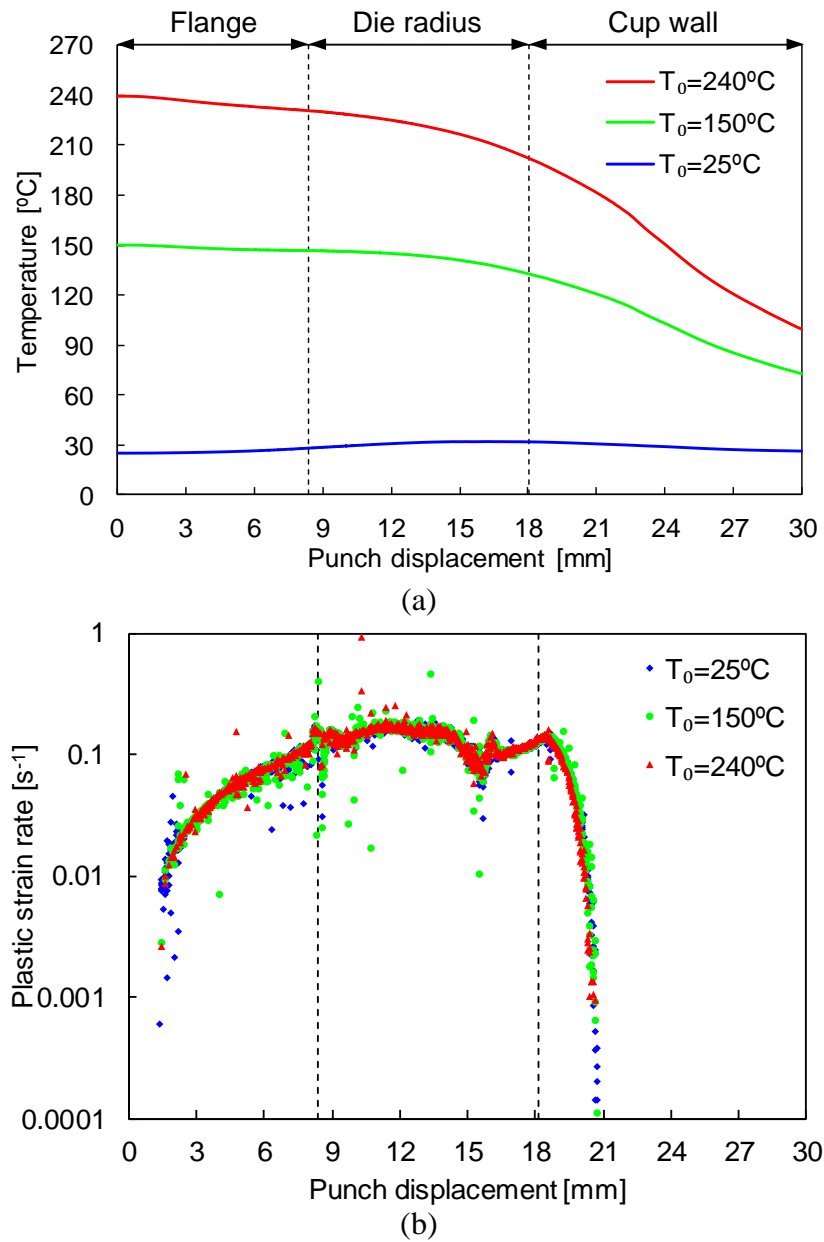

Figure 15: Analysis of the thermal conditions in the deep drawing of the cylindrical cup, considering a point initially located in the flange (5 mm from the perimeter): (a) temperature evolution; (b) evolution of the plastic strain rate. Results obtained with the Barlat'91 yield criterion.

Although the orthotropic behaviour of the blank changes with the temperature (see Table 3), the adopted yield criteria (Hill'48 and Barlat'91) are assumed temperature-independent. Therefore, the numerical simulations of the deep drawing process are performed using the different sets of parameters for the yield criterion (evaluated under isothermal conditions and listed in Table 5 and Table 7) according to the thermal conditions. The temperature of the blank at the beginning of the forming process (25ºC, 150ºC or 240ºC) is used to select the correct set of parameters. The results in Figure 15 (b) indicate that most of the plastic strain occurs while the material is located in the flange and die radius, where the temperature is roughly constant and similar to the one imposed to the die/blank-holder (see Figure 15 (a)). This indicates that the assumption adopted for describing the orthotropic behaviour of the material can provide a good description of the forming conditions.

## 5. Results and discussion

This section contains the comparison between experimental and numerical results of the warm deep drawing process, proposed as benchmark in the conference Numisheet 2016 (Manach et al., 2016). The main parameters evaluated are the evolution of the punch force during the forming operation, the thickness distribution measured in several directions and the final earing profile of the cup. The influence of the warm forming conditions on the springback is assessed by measuring the opening of a ring cut from the sidewall of the cylindrical cup.

### 5.1. Punch force

The comparison between experimental and numerical punch force evolution is presented in Figure 16 for the three different temperature conditions. For all process conditions analysed, the finite element solution is in very good agreement with the experimental results, particularly up to the instant the maximum force value is attained (approximately 11 mm of punch displacement). The punch force predicted by the von Mises and the Hill'48 yield criteria is slightly higher than the one predicted by the Barlat'91 (difference less than 10%). The stress state in the flange is pure compression at the outer radius and changes towards a shear state at the inner radius. Therefore, the difference in the punch force arises from the yield surface defined by the Barlat'91 yield criterion, which is globally located inside the ones defined for the other yield criteria in this domain, at least in the $\sigma_{11} - \sigma_{22}$ plane (see Figure 10). Since the temperature effect on the stress–strain curves up to 150ºC is very small (see Figure 7 for the higher value of strain rate), the punch force at room temperature and 150ºC are fairly similar. The sudden increase of the punch force, which is observed both experimentally and numerically at around 18 mm of displacement (Figure 16), is motived by the loss of contact between the blank and the blank-holder. After this instant, the experimental punch force is clearly overestimated by the numerical model, when the die/blank-holder are heated at 240ºC. This difference can be related with the underestimation of the blank temperature after this moment (see Figure 13), particularly for the material located near the die due to the loss of contact with the heated blank-holder, which induces a quick cooling of the blank (see Figure 15 (a)).

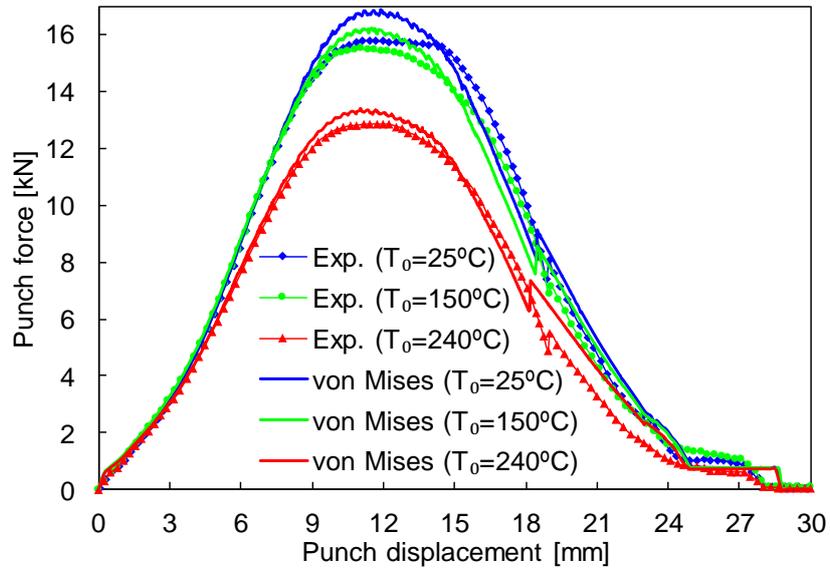

(a)

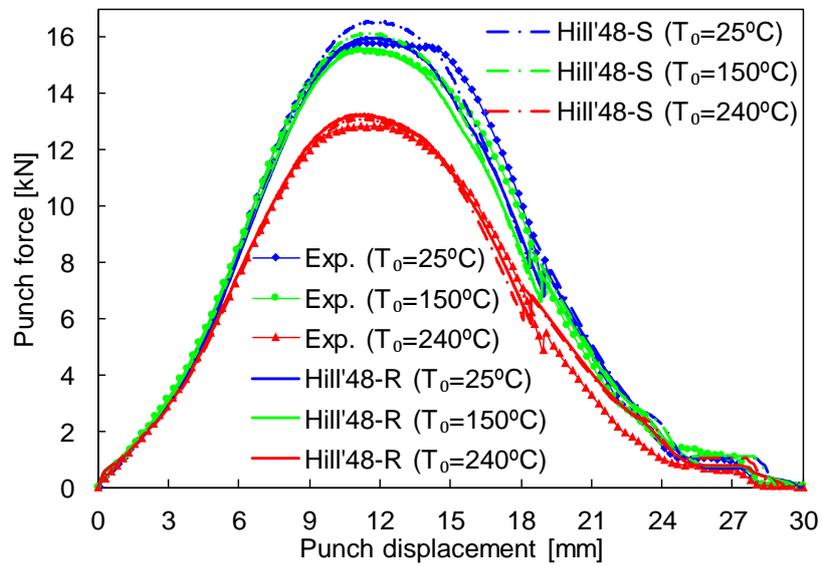

(b)

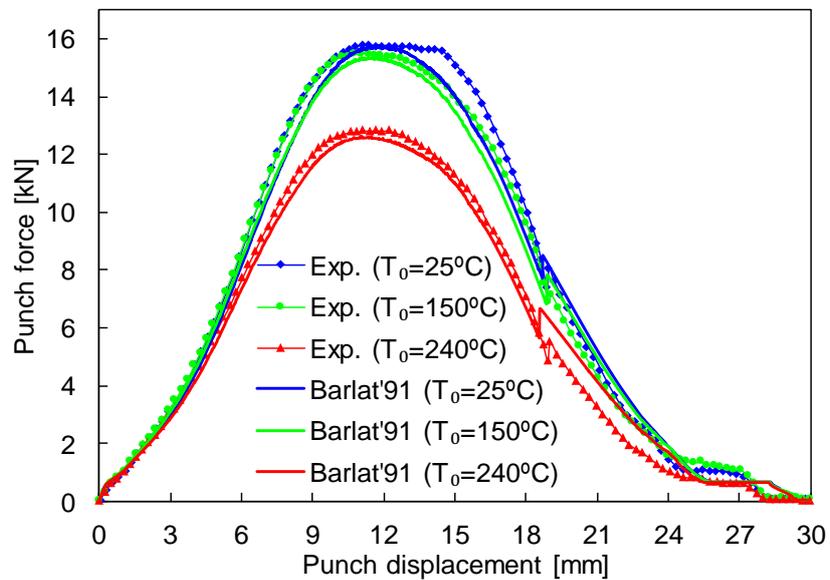

(c)

Figure 16: Comparison between experimental and numerical punch force evolution for three different temperature conditions using two yield functions: (a) von Mises; (b) Hill'48-R and Hill'48-S; (c) Barlat'91.

Since the blank-holder is allowed to move freely (absence of stopper), it establishes contact with the die when the flange of the cylindrical cup vanishes. Therefore, the cup rim (peak of the ears) is squeezed between the die and the blank-holder immediately before the blank loses contact with the blank-holder. The squeezing of the cup rim resulting from the warm deep drawing operation at 150ºC is presented in Figure 17, which compares the numerical with the experimental results. Since all cylindrical cups were experimentally trimmed to evaluate the springback using the split-ring test, only the ring containing the cup rim (see Figure 3 (a)) is presented. The large value of equivalent plastic strain predicted by the numerical simulation in the peak of the ears is in very good agreement with the experimental geometry of the interior cup rim, as shown in Figure 17. The sudden increase of the punch force shown in Figure 16 indicates the instant for which the squeezing of the ears is completed. Therefore, the squeezing of the cup rim occurs for all temperature conditions considered in the forming operation, which is accurately predicted by the finite element model.

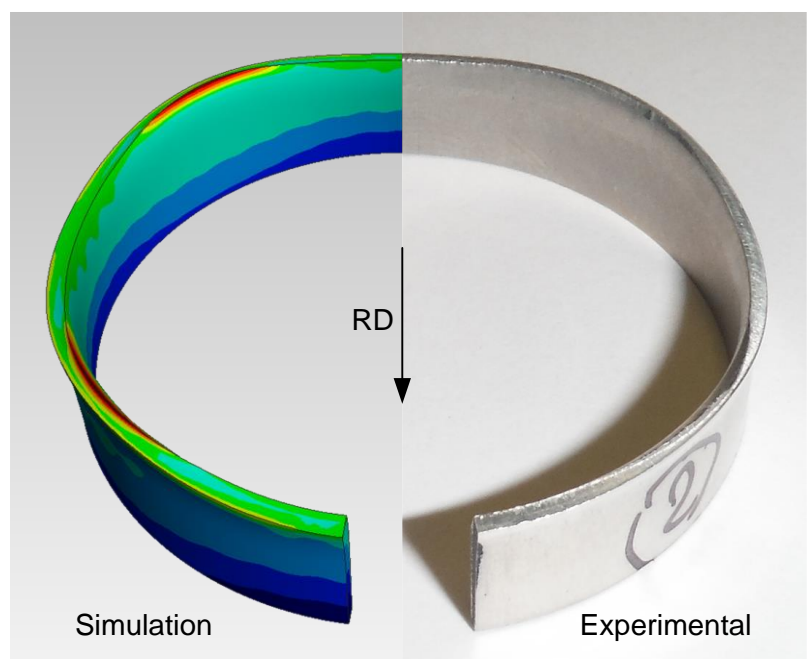

Figure 17: Squeezing of the cup rim during the drawing caused by the absence of blank-holder stopper. Warm forming at 150ºC (die and blank-holder) using the Barlat'91 yield criterion (contour plot of the equivalent plastic strain).

## 5.2. Thickness

The comparison between experimental and numerical thickness distribution in the cylindrical cup, predicted by the von Mises and Hill'48 yield criteria and evaluated in three directions (RD, DD and TD) is shown in Figure 18, for the three different temperature conditions. For all process conditions analysed, the zone of the cylindrical cup with a lower thickness value is located in the punch radius, while the higher thickness value occurs in the rim. Besides, the cup wall is always thinner in the DD due to the anisotropic behaviour of this aluminium alloy (see Figure 9), i.e. higher $r$-value in this direction. Globally, the thickness is underestimated in the bottom of the cup and overestimated in the cup wall, particularly when the anisotropy parameters of the Hill'48 yield criterion are evaluated using three yield stresses and one $r$-value (Hill'48-S). The comparison between the two numerical solutions obtained with the Hill'48 yield criterion allows concluding that the predicted thickness distribution is in better agreement with the experimental measurements when the anisotropy coefficients are evaluated using three $r$-values and one yield stress (Hill'48-R), as highlighted in Figure 18. For all heating conditions studied, the cup wall thickness predicted by the von Mises yield criterion is between the experimental values measured in the RD/TD and in the DD.

The final thickness of the cylindrical cup obtained under warm forming conditions (die and blank-holder at 150ºC) is similar to the one predicted considering the forming operation at room temperature, as shown in Figure 18. This behaviour was previously observed in the punch force (Figure 16), which results from the temperature-dependent flow stress and the weak dependence of the yield stresses with the temperature (Figure 9). On the other hand, the final thickness distribution increases about 0.015 mm when the forming operation is carried out at warm temperature (240ºC). However, the influence of the temperature conditions on the thickness is more pronounced in the bottom of the cup, which was also previously observed experimentally by (Kurukuri et al., 2009).

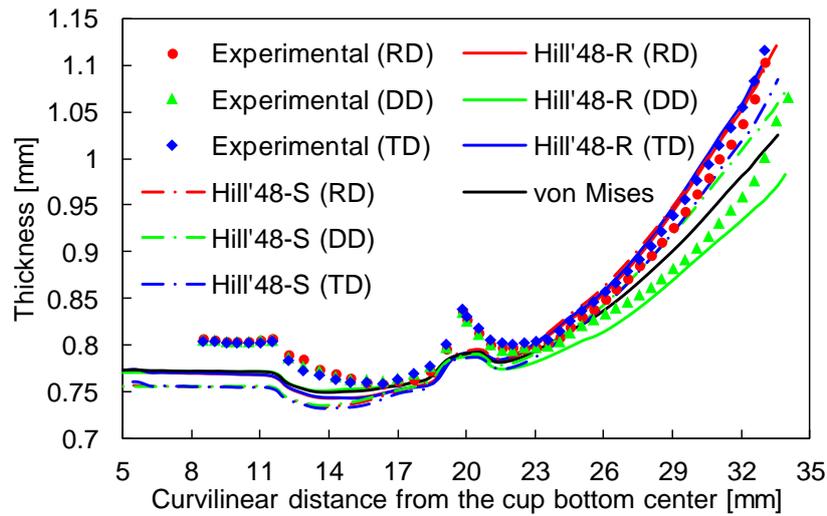

(a)

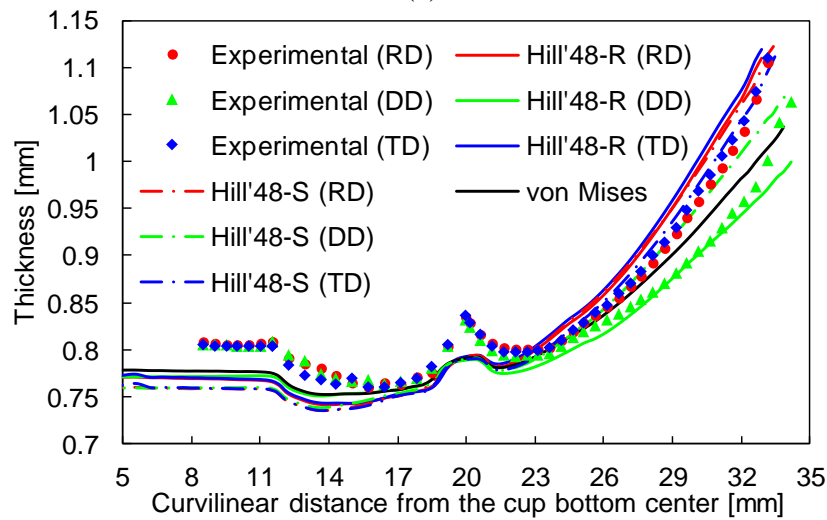

(b)

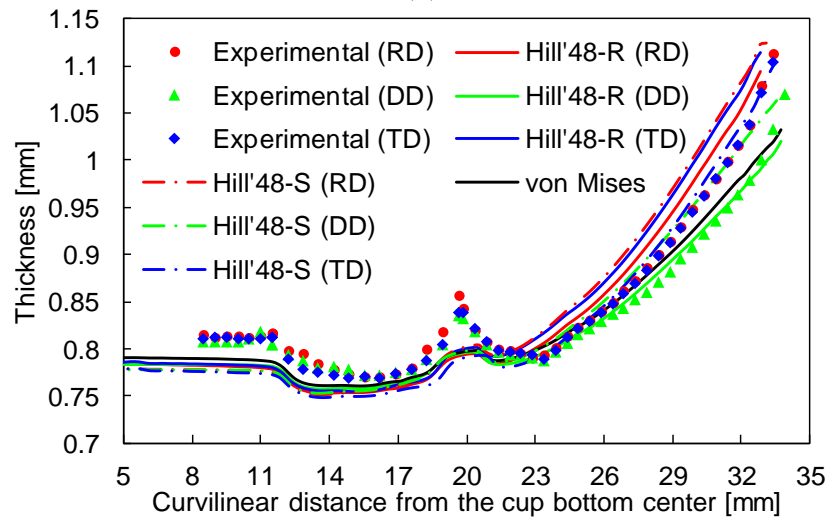

(c)

Figure 18: Comparison between experimental and numerical thickness distribution predicted by the von Mises and Hill'48 yield criteria, measured in three directions, for three different temperature conditions: (a) room temperature; (b) 150ºC; (c) 240ºC.

The comparison between experimental and numerical thickness distribution after forming, predicted by the Barlat'91 yield criterion and evaluated in three directions, is presented in Figure 19, for the three different temperature conditions. The final thickness is globally underestimated for all heating conditions, but particularly in the deep drawing operation at room temperature. Regarding the direction where the thickness is evaluated, the difference between experimental and numerical thickness is larger in the DD direction, as shown in Figure 19. However, the trend of the thickness distribution predicted by the numerical model is in good agreement with the experimental one, namely the variation of the cup wall thickness along the circumferential direction. The influence of the heating conditions on the final thickness distribution predicted by the Barlat'91 yield criterion is identical to the one obtained with the Hill'48 yield criterion. Nevertheless, the thickness in the bottom of the cup increases about 0.010 mm by performing the warm forming at 240ºC.

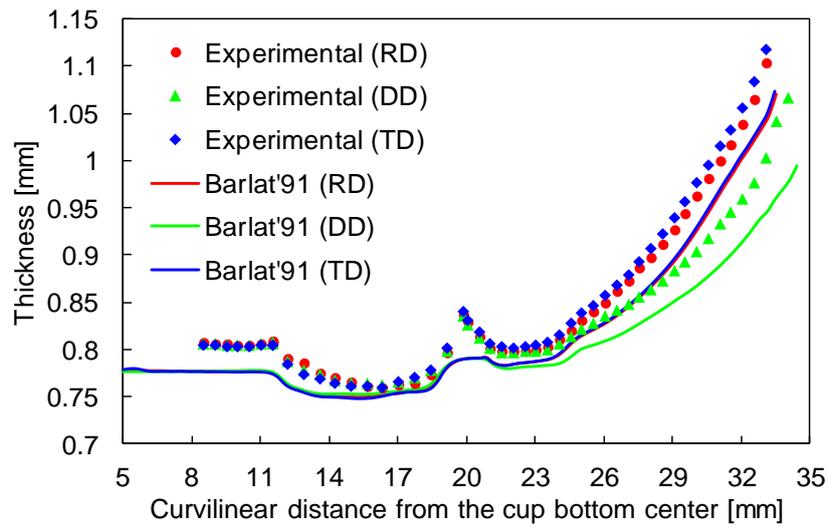

(a)

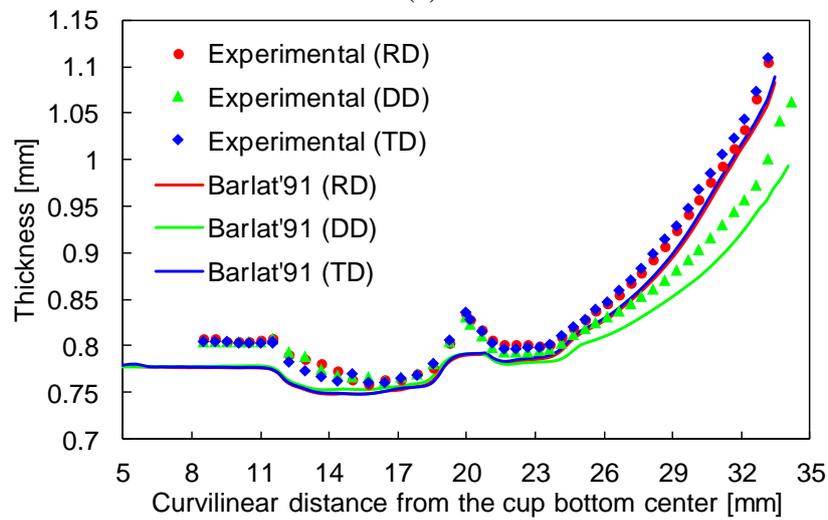

(b)

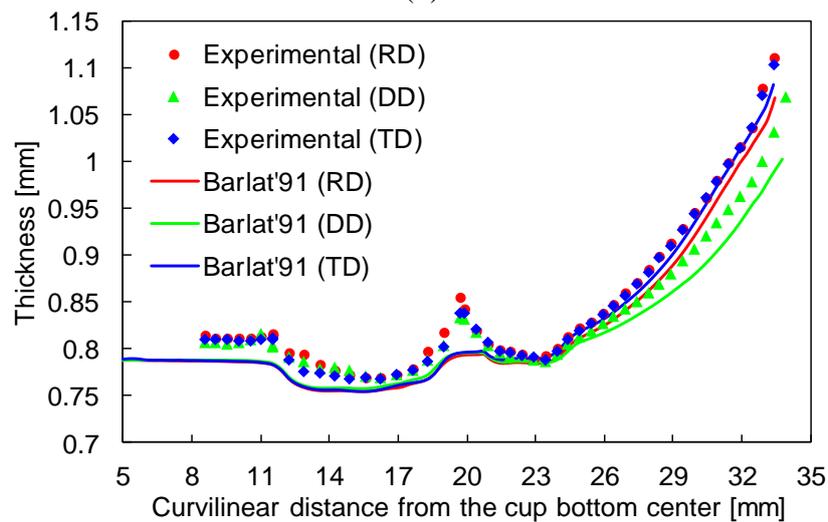

(c)

Figure 19: Comparison between experimental and numerical thickness distribution predicted by the Barlat'91 yield criterion, measured in three directions, for three different temperature conditions: (a) room temperature; (b) 150ºC; (c) 240ºC.

Since the thickness experimentally measured in the bottom of the cup is higher than the nominal thickness of the sheet (0.8 mm), probably the effective thickness of the sheet adopted in the experiments was larger than 0.8 mm. Accordingly, the finite element simulation of the deep drawing process was carried out using a blank with an initial thickness of 0.815 mm. Considering room temperature conditions, the comparison between the experimental and the numerical thickness distributions is presented in Figure 20, using both the Hill'48-R and Barlat'91 yield functions. Adopting the Hill'48 yield criterion, the cup wall thickness is globally overestimated while the thickness in the bottom of the cup is clearly underestimated. On the other hand, the experimental thickness distribution is only slightly underestimated by the Barlat'91 yield criterion, as shown in Figure 20 (b). Indeed, the trend of the thickness distribution predicted by the numerical model is in good agreement with the experimental one, for all directions.

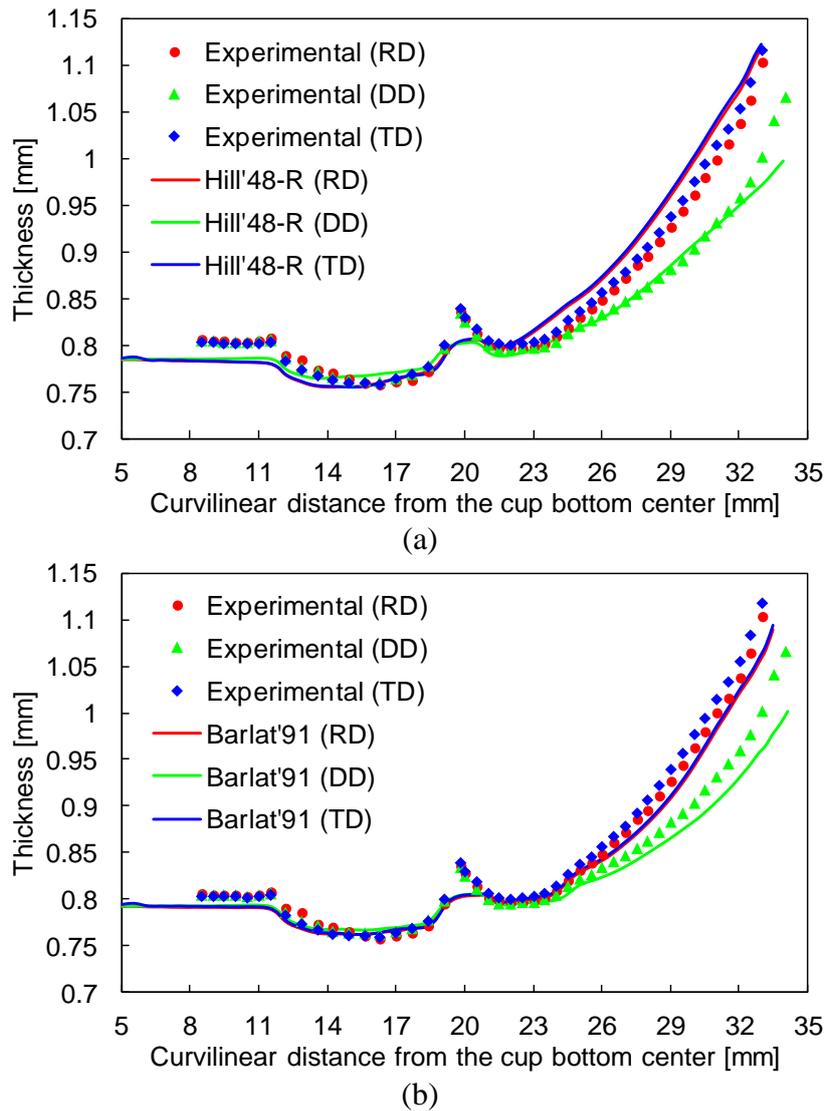

Figure 20: Comparison between experimental and numerical thickness distribution, measured in three directions, considering a blank thickness of 0.815 mm and room temperature conditions, using two yield functions: (a) Hill'48-R; (b) Barlat'91.

Regarding the warm forming at 240ºC and considering the blank with an initial thickness of 0.815 mm, the comparison between the experimental and the numerical thickness distributions is presented in Figure 21, comparing the yield functions Hill'48-R and Barlat'91. Adopting the Hill'48 yield criterion, the cup wall thickness is clearly overestimated in all directions, while the thickness in the bottom of the cup is underestimated. On the other hand, the thickness distribution predicted by the Barlat'91 yield criterion is in very good agreement with the experimental measurements, both in the wall and bottom of the cup (all directions), as shown in Figure 21 (b). According to Figure 20 and Figure 21, the cup thickness distribution can be accurately predicted by the Barlat'91 yield criterion, considering a slight increase of the blank initial thickness. Moreover, the punch force increases about 2% around the maximum value, improving the accuracy of the numerical solution (see Figure 16 (c)).

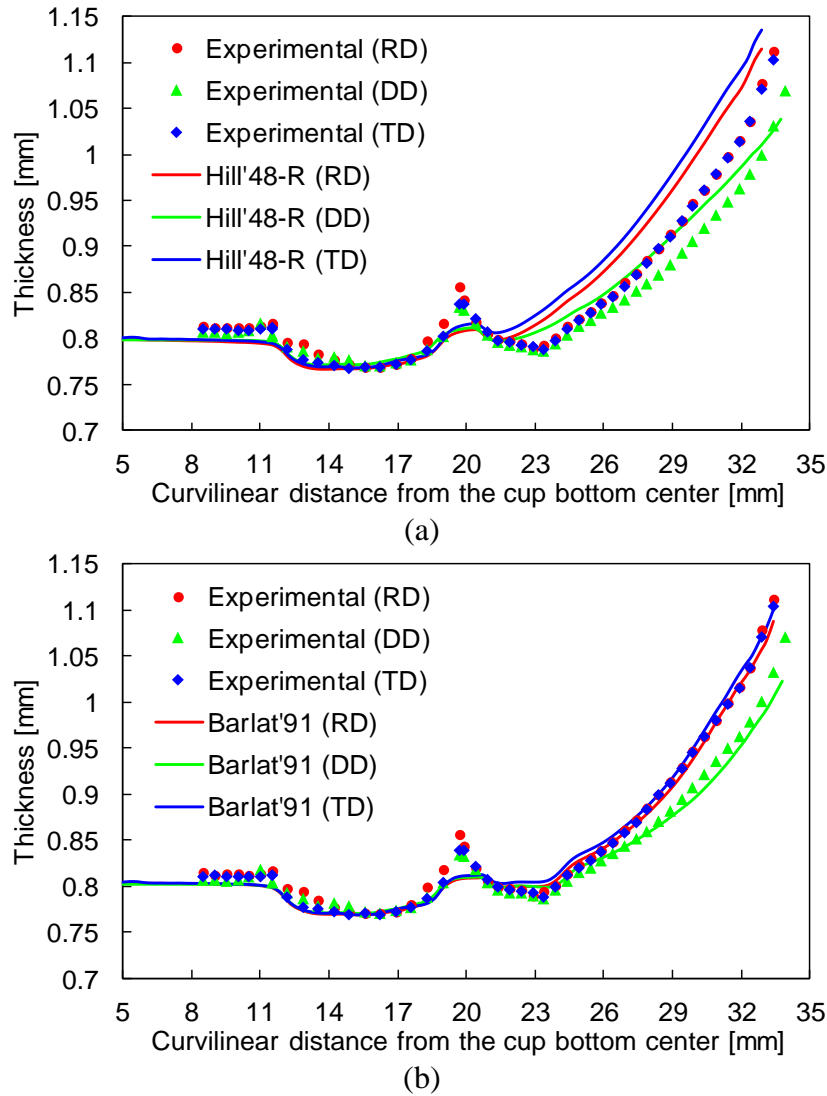

Figure 21: Comparison between experimental and numerical thickness distribution, measured in three directions, considering a blank thickness of 0.815 mm and heated die/blank-holder up to 240ºC, using two yield functions: (a) Hill'48-R; (b) Barlat'91.

### 5.3. Earing profile

The ears generated during the deep drawing of an AA5754-O aluminium cylindrical cup are directly related with the anisotropic behaviour of the sheet (Figure 9). Recent studies indicate that the accurate prediction of the earing profile requires the simultaneous description of both the *r*-value and yield stress directionalities (Chung et al., 2011; Yoon et al., 2011). The comparison between experimental and numerical earing profile is presented in Figure 22, for the three different temperature conditions. Both anisotropic yield criteria selected (Hill'48 and Barlat'91) predict four ears in the cup drawing, which is in accordance with the experiments (Manach et al., 2016). Nevertheless, globally the experimental cup height is underestimated by the numerical simulation. For all temperature conditions considered, the earing profile presents the maxima at RD and TD positions and minimum at DD and equivalent positions, which is consistent with the *r*-values distribution (see Figure 9 (a)). The amplitude of the ears predicted by the Hill'48 yield criterion is strongly influenced by the data used in the evaluation of the anisotropy parameters (Neto et al., 2014a). Since the Hill'48-S model presents both yield stresses and the *r*-values with lower variations in the plane of the sheet (see Figure 9), the amplitude of the ears predicted by the Hill'48-S model is substantially lower, as shown in Figure 22 (b). The amplitude of the ears predicted by the Barlat'91 yield criterion is in-between the values predicted by the two sets of parameters identified for the Hill'48 yield criterion, as also observed in other studies (Yoon et al., 1999). Moreover, the average cup height is about 0.3 mm larger using the Barlat'91 yield criterion, which is related with the sharper shape of the yield surface (see Figure 10). According to the preceding analysis, it should be mentioned that a slight increase of the initial blank thickness leads to a negligible influence on the earing profile and cup height.

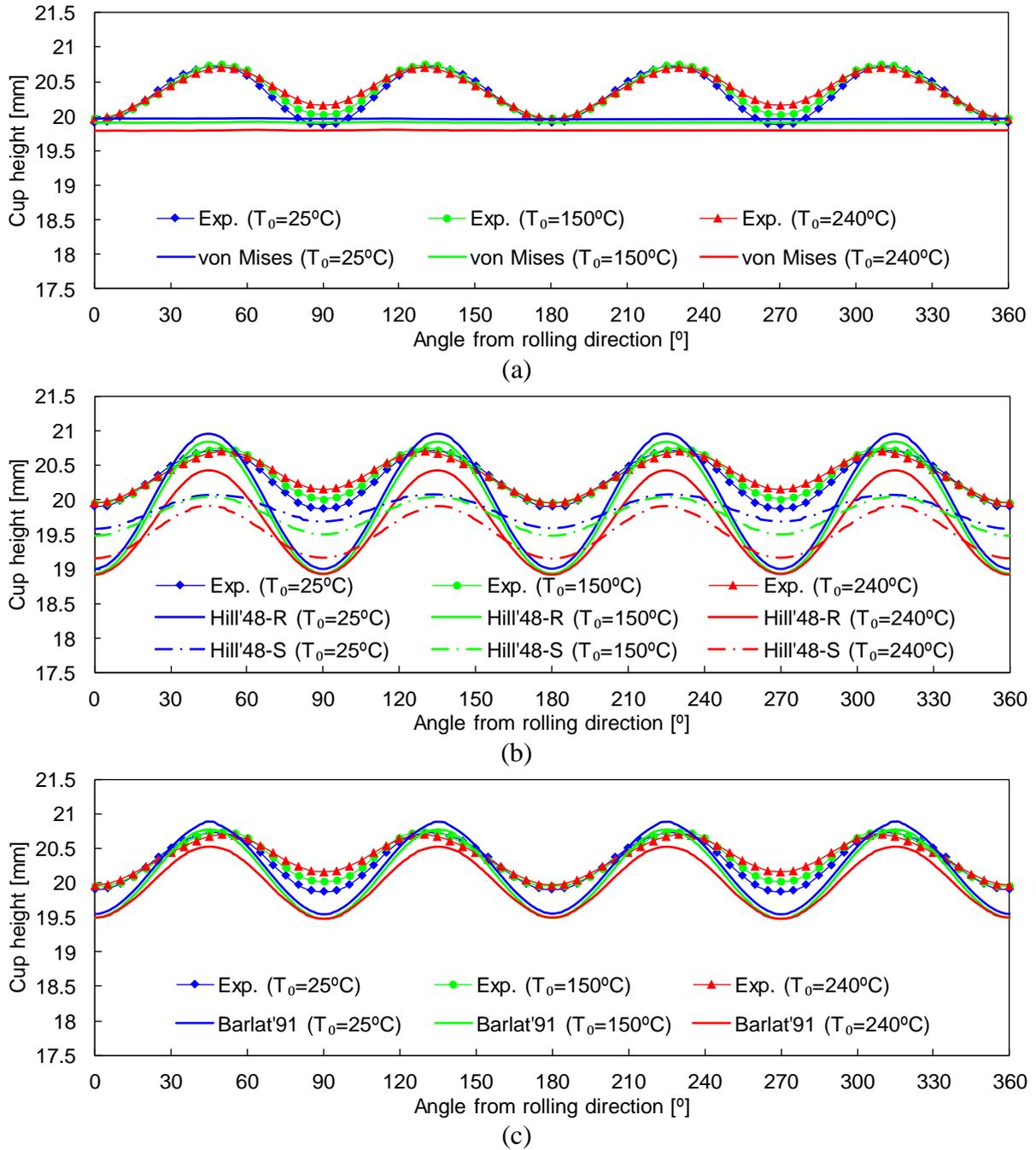

Figure 22: Comparison between experimental and numerical earing profile after cup forming for three different temperature conditions and using two yield functions: (a) von Mises; (b) Hill'48-R and Hill'48-S; (c) Barlat'91.

Except when adopting the Hill'48-S model, the amplitude of the ears decreases with the temperature rise. This is particularly evident in the warm forming at 240ºC (die and blank-holder), where the temperature effect on the flow stress is more pronounced (see Figure 7). The experimental results presented by (Ghosh et al., 2014) on a 6xxx aluminium alloy show that the number of ears does not change with the temperature rise but their amplitude is reduced. In terms of predicted earing profile, the temperature effect on the yield function is more important than on the hardening law. In fact, the distribution of the yield stress (see Figure 9 (b)) together with the *r*-values distribution present a strong impact on the amplitude of the ears. Numerical results show that increasing the test temperature up to 240ºC but using the anisotropic behaviour at room temperature, leads to a reduction of the cup height of about 0.01 mm, while the earing trend is identical. Except for the Hill'48-S model, the influence of the process temperature conditions on the earing profile predicted by numerical simulation is most pronounced in the DD position, while experimentally is most evident in the TD location, as shown in Figure 22. Since the earing profile around TD is predominantly dictated by the *r*-value and the yield stress nearby the RD (Chung et al., 2011; Yoon et al., 2011), accordingly to Figure 9 is not expected to obtain numerically significant differences with the temperature rise.

### 5.4. Springback

The effect of the warm forming temperature on the springback is evaluated by the split-ring test, previously described in Section 2.2. The finite element analysis of the split-ring test requires a preceding stage of cup cooling down to room temperature, for the tests performed at warm conditions. After that, the ring is cut from the sidewall of the cylindrical cup, as illustrated in Figure 3. This numerical procedure is carried out with the in-house code DD3TRIM. The geometrical trimming of the finite element mesh (cup) is performed using the algorithm proposed by (Baptista et al., 2006), while the remapping of the state variables to the new mesh (ring) is performed with the incremental Remapping Methods (IVR) method described in (Neto et al., 2016). The last stage comprises the split of the ring and consequent springback evaluation.

The comparison between the experimental and numerical values of ring opening is presented in Figure 23 for the three different temperature conditions, highlighting the strong impact of the yield criterion (von Mises, Hill'48-R, Hill'48-S and Barlat'91) on the springback value. In fact, the experimental ring opening is underestimated by all yield criteria for all heating conditions. Nevertheless, all numerical models predict the reduction of the springback with the temperature rise, as shown in Figure 23. This trend was also numerically predicted by (Kim and Koç, 2008) in a simple draw bending process, which is associated with the decreased material strength at elevated temperatures. The ring opening predicted by the numerical model using the isotropic yield criterion (von Mises) is in better agreement with the experimental one, particularly at room temperature and at 240ºC, where the difference is inferior to 0.3 mm. On the other hand, the finite element simulations using the Hill'48-S yield criterion provide the worst approximation for the springback prediction, with an underestimation of the ring opening of more than 45% in warm forming conditions (see Figure 23). The ring opening predicted by the Barlat'91 yield criterion is in-between the values predicted by the Hill'48 yield criterion, as shown in Figure 23. Considering the warm forming at 240ºC, the slight increase of the blank thickness to 0.815 mm (see Section 5.2) leads to a reduction of the predicted ring opening of about 2.5%. These results concerning the influence of the yield criterion are in agreement with the previous studies (Laurent et al., 2015, 2009), where the ring opening is clearly underestimated (at least 35%) using the Hill'48 yield criterion. Previous studies shown that the ring opening is underestimated by DD3IMP (Laurent et al., 2010).

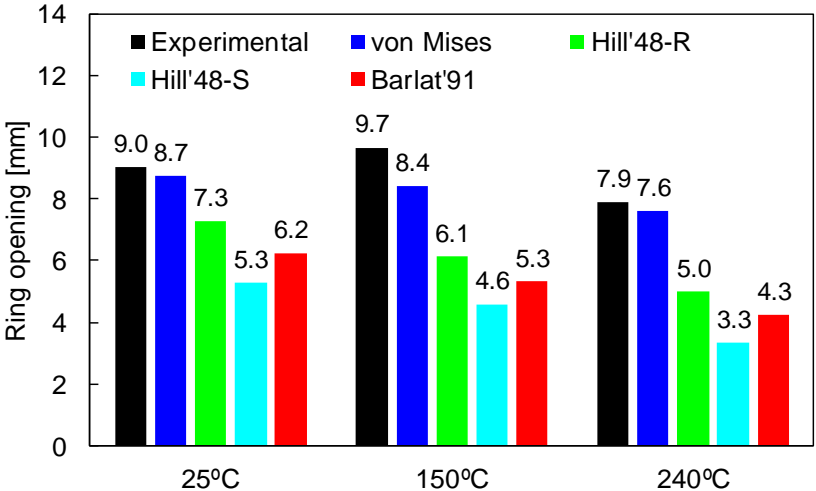

Figure 23: Comparison between experimental and numerical values of ring opening for the three different temperature conditions and using three yield functions (von Mises, Hill'48-R, Hill'48-S and Barlat'91).

Since the value of the ring opening is dictated by the residual stress state in the ring before splitting, the distribution of the hoop stress is analysed, which is the main factor influencing the springback (Laurent et al., 2010). The bending moment given by the integration of the hoop stress over the ring thickness defines the shape of the ring after springback and consequently the opening value (Simões et al., 2017). However, the hoop stress distribution varies along the circumferential direction of the ring due to the anisotropic behaviour of the material. Considering the deep drawing at room temperature, the predicted distribution of the hoop stress on the ring (before splitting) is presented in Figure 24, comparing the various yield functions (von Mises, Hill'48-R, Hill'48-S and Barlat'91). The predicted hoop stress is compressive in the inner surface of the ring and tensile on the outer surface. Besides, the distribution predicted by the numerical model that considers material isotropy (von Mises) is uniform along the circumferential direction, as shown in Figure 24 (a). On the other hand, adopting an anisotropic yield criterion (Hill'48 and Barlat'91), the predicted hoop stress distribution presents a variation along the circumferential direction. This stress gradient is more pronounced using the Hill'48-R yield criterion (Figure 24), which is in accordance with the earing profile shown in Figure 22. Therefore, the low value of the ring opening predicted by the Hill'48-S yield criterion (see Figure 23) is caused by the lower hoop stress gradient through the thickness integrated over the circumferential direction.

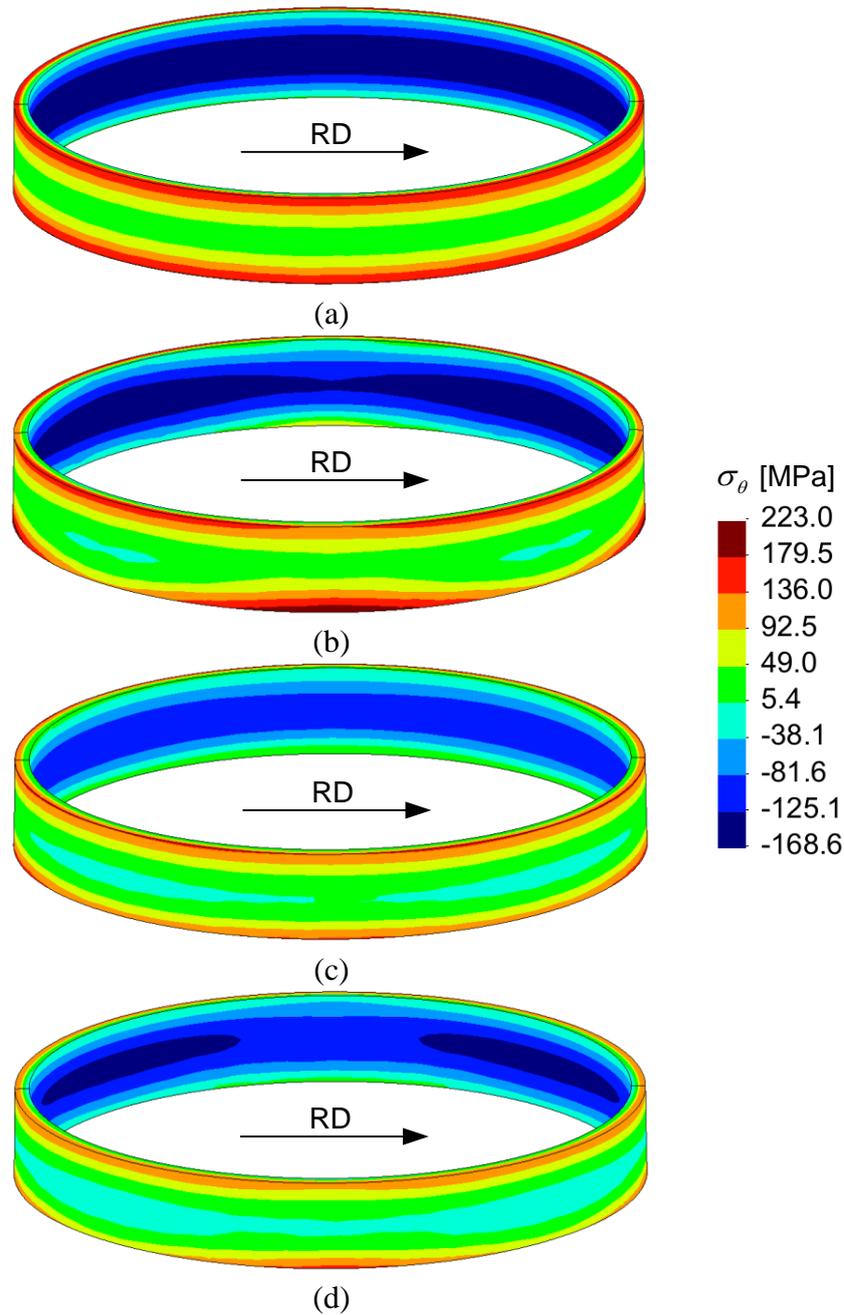

Figure 24: Numerical distribution of the hoop stress on the ring before splitting (deep drawing at room temperature) using three yield functions: (a) von Mises; (b) Hill'48-R; (c) Hill'48-S; (d) Barlat'91.

For all temperature conditions considered, the distribution of the hoop stress in the cross section of the ring aligned with the RD is presented in Figure 25, comparing the various yield criteria (von Mises, Hill'48-R and Barlat'91). The hoop stress gradient through the thickness decreases with the temperature rise due to the material softening (Figure 7), which is in accordance with the ring opening values presented in Figure 23. The variation of the hoop stress along the vertical direction (ring height) is mainly caused by the release of residual stresses resulting from the ring cutting operation (before split). This gradient is more pronounced when the isotropic yield criterion is adopted, as shown in Figure 25.

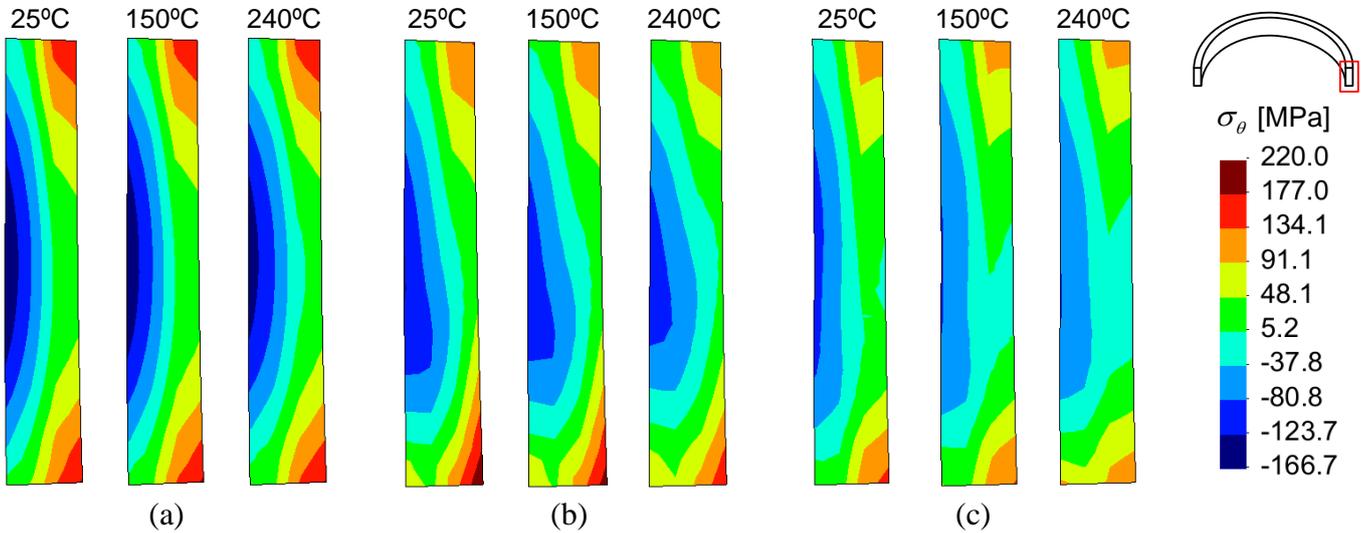

Figure 25: Numerical distribution of the hoop stress in the cross section of the ring aligned with the RD before splitting for three different temperature conditions and using various yield functions: (a) von Mises; (b) Hill'48-R; (c) Barlat'91.

As previously mentioned, experimental results show the reduction of the Young modulus with the temperature rise (Laurent et al., 2015; Lee et al., 2015). On the other hand, the Young modulus degradation with accumulated plastic strain has been experimentally observed for several metals (Cleveland and Ghosh, 2002). Nevertheless, the springback results presented in Figure 23 were predicted using a constant value of Young modulus ($E$=71.7 GPa) in the numerical model. Therefore, it is expected that the ring opening prediction can be improved taking into account both the Young modulus degradation and its evolution with the temperature.

Due to the lack of experimental data regarding the aluminium alloy under analysis, the influence of the Young modulus on the ring opening is numerically assessed at room temperature. Hence a reduced constant value of Young modulus ($E$=55.0 GPa) is used in the simulation of the cup forming at room temperature, comparing three yield functions (von Mises, Hill'48-R and Barlat'91). The predicted ring opening is presented in Figure 26, using the two values of Young modulus, for comparison. The strong impact of the Young modulus value on the springback is highlighted, i.e. the predicted ring opening increases at least 40% when the Young modulus is reduced about 20%. This trend is similar for all yield criteria considered, as shown in Figure 26. However, it should be mentioned that the reduction of the Young modulus presents a negligible effect on all other process variables (punch force, thickness distribution and earing profile).

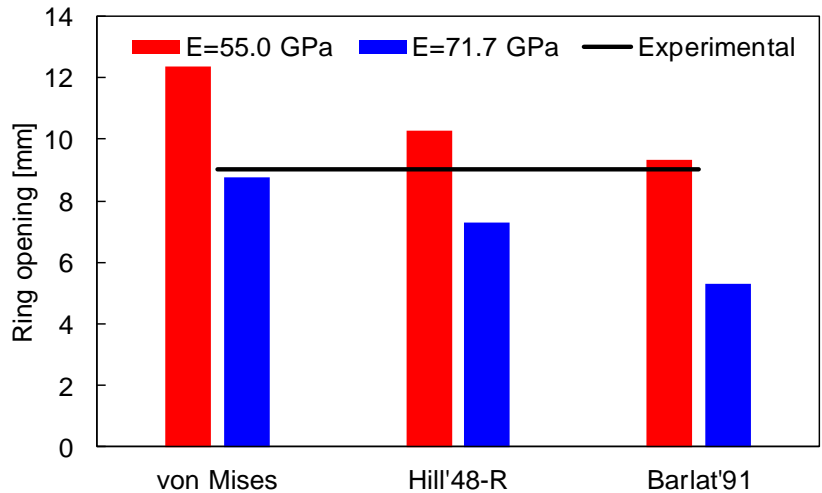

Figure 26: Influence of the Young modulus (constant) on the predicted ring opening at room temperature forming conditions, comparing three yield functions (von Mises, Hill'48-R and Barlat'91).

## 6. Conclusions

This study presents the deep drawing simulation of a cylindrical cup, performed both at room temperature and in warm conditions (heated die/blank-holder and cooled punch). The effect of the heating conditions on the springback is evaluated by means of the split-ring test, where a ring is cut from the sidewall of the cup. Based on the analysis of the forming conditions, the coupled thermo-mechanical finite element analysis is carried out using a rate-independent thermo-elasto-plastic material model for the blank (AA5086 aluminium alloy). The parameters of the hardening law are identified using data from uniaxial tensile tests performed at a strain rate of 0.1 s$^{-1}$, at temperatures from 25ºC to 240ºC. On the other hand, the yield function is assumed temperature-independent but the set of anisotropy parameters is identified for three different temperatures, according to the forming conditions. Regarding the transient heat conduction problem, the heat transfer between the forming tools and the blank is defined by the interfacial heat transfer coefficient, which takes into account the gap distance between the bodies.

Concerning the forming operation, the evolution of the punch force, the final thickness distribution in several directions and the earing profile of the cup are the main parameters analysed. The predicted punch force is predominantly dictated by the accurate description of the flow stress at different temperatures, which is defined by the temperature-dependent hardening law. Thus, the punch force decreases with the temperature rise, particularly at 240ºC. The predicted thickness distribution is clearly overestimated in the cup wall using the Hill'48 yield criterion, while the adoption of the Barlat'91 yield criterion leads to numerical results in very good agreement with the experimental measurements. Nevertheless, the predicted thickness is only slightly influenced by the heating conditions used in the deep drawing operation. The cup height is globally underestimated by the numerical model but the earing profile (four ears) is accurately predicted. Indeed, the amplitude of the ears predicted by the Hill'48-R yield criterion is substantially higher than the one obtained using the Barlat'91 yield criterion, which is in better agreement with the experimental one.

For all models studied, increasing the forming temperature leads to a reduction of the springback value. However, the predicted springback, evaluated through the ring opening, is strongly influenced by the yield function adopted to model the material anisotropy. In fact, considering a constant Young modulus value of $E$=71.7 GPa, the numerical predictions obtained with the von Mises yield criterion are in better agreement with the experimental values. However, for all yield criteria under analysis, considering room temperature conditions, the predicted ring opening increases at least 40% when the constant value for the Young modulus is 20% lower, highlighting the importance of an accurate modelling of the elastic behaviour of the blank.

## Acknowledgements

The authors gratefully acknowledge the financial support of the Portuguese Foundation for Science and Technology (FCT) under projects PTDC/EMS-TEC/0702/2014 (POCI-01-0145-FEDER-016779) and PTDC/EMS-TEC/6400/2014 (POCI-01-0145-FEDER-016876) by UE/FEDER through the program COMPETE 2020. The first author is also grateful to the FCT for the Postdoctoral grant SFRH/BPD/101334/2014. We also would like to acknowledge the benchmark committee to make available the experimental data used in the present study.

# References

Abedrabbo, N., Pourboghrat, F., Carsley, J., 2007. Forming of AA5182-O and AA5754-O at elevated temperatures using coupled thermo-mechanical finite element models. Int. J. Plast. 23, 841–875. doi:10.1016/j.ijplas.2006.10.005

Abedrabbo, N., Pourboghrat, F., Carsley, J., 2006. Forming of aluminum alloys at elevated temperatures – Part 1: Material characterization. Int. J. Plast. 22, 314–341. doi:10.1016/j.ijplas.2005.03.005

Adam, L., Ponthot, J.-P., 2005. Thermomechanical modeling of metals at finite strains: First and mixed order finite elements. Int. J. Solids Struct. 42, 5615–5655. doi:10.1016/j.ijsolstr.2005.03.020

Alghtani, A.H., 2015. Analysis and Optimization of Springback in Sheet Metal Forming. University of Leeds.

Argyris, J.H., Doltsinis, J.S., 1981. On the natural formulation and analysis of large deformation coupled thermomechanical problems. Comput. Methods Appl. Mech. Eng. 25, 195–253. doi:10.1016/0045-7825(81)90084-0

Armero, F., Simo, J.C., 1993. A priori stability estimates and unconditionally stable product formula algorithms for nonlinear coupled thermoplasticity. Int. J. Plast. 9, 749–782. doi:10.1016/0749-6419(93)90036-P

Armero, F., Simo, J.C., 1992. A new unconditionally stable fractional step method for non-linear coupled thermomechanical problems. Int. J. Numer. Methods Eng. 35, 737–766. doi:10.1002/nme.1620350408

Baptista, A.J., Alves, J.L., Rodrigues, D.M., Menezes, L.F., 2006. Trimming of 3D solid finite element meshes using parametric surfaces: Application to sheet metal forming. Finite Elem. Anal. Des. 42, 1053–1060. doi:10.1016/j.finel.2006.03.005

Barlat, F., Aretz, H., Yoon, J.W., Karabin, M.E., Brem, J.C., Dick, R.E., 2005. Linear transfomation-based anisotropic yield functions. Int. J. Plast. 21, 1009–1039. doi:10.1016/j.ijplas.2004.06.004

Barlat, F., Lege, D.J., Brem, J.C., 1991. A six-component yield function for anisotropic materials. Int. J. Plast. 7, 693–712. doi:10.1016/0749-6419(91)90052-Z

Barros, P.D., Carvalho, P.D., Alves, J.L., Oliveira, M.C., Menezes, L.F., 2016. DD3MAT - a code for yield criteria anisotropy parameters identification. J. Phys. Conf. Ser. 734, 32053. doi:10.1088/1742-6596/734/3/032053

Berisha, B., Hora, P., Wahlen, A., Tong, L., 2010. A combined isotropic-kinematic hardening model for the simulation of warm forming and subsequent loading at room temperature. Int. J. Plast. 26, 126–140. doi:10.1016/j.ijplas.2009.06.001

Bernard, C., Coër, J., Laurent, H., Manach, P.Y., Oliveira, M., Menezes, L.F., 2016. Influence of Portevin-Le Chatelier Effect on Shear Strain Path Reversal in an Al-Mg Alloy at Room and High Temperatures. Exp. Mech. 1–11. doi:10.1007/s11340-016-0229-z

Bolt, P.., Lamboo, N.A.P.., Rozier, P.J.C.., 2001. Feasibility of warm drawing of aluminium products. J. Mater. Process. Technol. 115, 118–121. doi:10.1016/S0924-0136(01)00743-9

Caron, E., Daun, K.J., Wells, M.A., 2013. Experimental Characterization of Heat Transfer

Coefficients During Hot Forming Die Quenching of Boron Steel. Metall. Mater. Trans. B 44, 332–343. doi:10.1007/s11663-012-9772-x

Caron, E.J.F.R., Daun, K.J., Wells, M.A., 2014. Experimental heat transfer coefficient measurements during hot forming die quenching of boron steel at high temperatures. Int. J. Heat Mass Transf. 71, 396–404. doi:10.1016/j.ijheatmasstransfer.2013.12.039

Chen, P., Chen, S., 2013. Thermo-mechanical contact behavior of a finite graded layer under a sliding punch with heat generation. Int. J. Solids Struct. 50, 1108–1119. doi:10.1016/j.ijsolstr.2012.12.007

Chung, K., Kim, D., Park, T., 2011. Analytical derivation of earing in circular cup drawing based on simple tension properties. Eur. J. Mech. - A/Solids 30, 275–280. doi:10.1016/j.euromechsol.2011.01.006

Cleveland, R.., Ghosh, A.., 2002. Inelastic effects on springback in metals. Int. J. Plast. 18, 769–785. doi:10.1016/S0749-6419(01)00054-7

Coër, J., Bernard, C., Laurent, H., Andrade-Campos, A., Thuillier, S., 2011. The Effect of Temperature on Anisotropy Properties of an Aluminium Alloy. Exp. Mech. 51, 1185–1195. doi:10.1007/s11340-010-9415-6

Coër, J., Manach, P.Y., Laurent, H., Oliveira, M.C., Menezes, L.F., 2013. Piobert–Lüders plateau and Portevin–Le Chatelier effect in an Al–Mg alloy in simple shear. Mech. Res. Commun. 48, 1–7. doi:10.1016/j.mechrescom.2012.11.008

Dasappa, P., Inal, K., Mishra, R., 2012. The effects of anisotropic yield functions and their material parameters on prediction of forming limit diagrams. Int. J. Solids Struct. 49, 3528–3550. doi:10.1016/j.ijsolstr.2012.04.021

de Codes, R.N., Hopperstad, O.S., Engler, O., Lademo, O.-G., Embury, J.D., Benallal, A., 2011. Spatial and Temporal Characteristics of Propagating Deformation Bands in AA5182 Alloy at Room Temperature. Metall. Mater. Trans. A 42, 3358–3369. doi:10.1007/s11661-011-0749-1

Demeri, M.Y., Lou, M., Saran, M.J., 2000. A Benchmark Test for Springback Simulation in Sheet Metal Forming. doi:10.4271/2000-01-2657

Erbts, P., Düster, A., 2012. Accelerated staggered coupling schemes for problems of thermoelasticity at finite strains. Comput. Math. with Appl. 64, 2408–2430. doi:10.1016/j.camwa.2012.05.010

Ghosh, M., Miroux, A., Werkhoven, R.J., Bolt, P.J., Kestens, L.A.I., 2014. Warm deep-drawing and post drawing analysis of two Al–Mg–Si alloys. J. Mater. Process. Technol. 214, 756–766. doi:10.1016/j.jmatprotec.2013.10.020

Grèze, R., Manach, P.Y., Laurent, H., Thuillier, S., Menezes, L.F., 2010. Influence of the temperature on residual stresses and springback effect in an aluminium alloy. Int. J. Mech. Sci. 52, 1094–1100. doi:10.1016/j.ijmecsci.2010.04.008

Hill, R., 1948. A Theory of the Yielding and Plastic Flow of Anisotropic Metals. Proc. R. Soc. A Math. Phys. Eng. Sci. 193, 281–297. doi:10.1098/rspa.1948.0045

Hippke, H., Manopulo, N., Yoon, J.W., Hora, P., 2017. On the efficiency and accuracy of stress integration algorithms for constitutive models based on non-associated flow rule. Int. J. Mater. Form. 1–8. doi:10.1007/s12289-017-1347-6

Hirsch, J., 2011. Aluminium in Innovative Light-Weight Car Design. Mater. Trans. 52, 818–

824. doi:10.2320/matertrans.L-MZ201132

Hockett, J.E., Sherby, O.D., 1975. Large strain deformation of polycrystalline metals at low homologous temperatures. J. Mech. Phys. Solids 23, 87–98. doi:10.1016/0022-5096(75)90018-6

Hosford, W.F., 1972. A Generalized Isotropic Yield Criterion. J. Appl. Mech. 39, 607. doi:10.1115/1.3422732

Hughes, T.J.R., 1980. Generalization of selective integration procedures to anisotropic and nonlinear media. Int. J. Numer. Methods Eng. 15, 1413–1418. doi:10.1002/nme.1620150914

Kabirian, F., Khan, A.S., Pandey, A., 2014. Negative to positive strain rate sensitivity in 5xxx series aluminum alloys: Experiment and constitutive modeling. Int. J. Plast. 55, 232–246. doi:10.1016/j.ijplas.2013.11.001

Kim, H.S., Koç, M., 2008. Numerical investigations on springback characteristics of aluminum sheet metal alloys in warm forming conditions. J. Mater. Process. Technol. 204, 370–383. doi:10.1016/j.jmatprotec.2007.11.059

Kim, H.S., Koç, M., Ni, J., Ghosh, A., 2006. Finite Element Modeling and Analysis of Warm Forming of Aluminum Alloys—Validation Through Comparisons With Experiments and Determination of a Failure Criterion. J. Manuf. Sci. Eng. 128, 613. doi:10.1115/1.2194065

Kim, S., Lee, J., Barlat, F., Lee, M.-G., 2013. Formability prediction of advanced high strength steels using constitutive models characterized by uniaxial and biaxial experiments. J. Mater. Process. Technol. 213, 1929–1942. doi:10.1016/j.jmatprotec.2013.05.015

Kurukuri, S., van den Boogaard, A.H., Miroux, A., Holmedal, B., 2009. Warm forming simulation of Al–Mg sheet. J. Mater. Process. Technol. 209, 5636–5645. doi:10.1016/j.jmatprotec.2009.05.024

Laurent, H., Coër, J., Manach, P.Y., Oliveira, M.C., Menezes, L.F., 2015. Experimental and numerical studies on the warm deep drawing of an Al–Mg alloy. Int. J. Mech. Sci. 93, 59–72. doi:10.1016/j.ijmecsci.2015.01.009

Laurent, H., Grèze, R., Manach, P.Y., Thuillier, S., 2009. Influence of constitutive model in springback prediction using the split-ring test. Int. J. Mech. Sci. 51, 233–245. doi:10.1016/j.ijmecsci.2008.12.010

Laurent, H., Grèze, R., Oliveira, M.C., Menezes, L.F., Manach, P.Y., Alves, J.L., 2010. Numerical study of springback using the split-ring test for an AA5754 aluminum alloy. Finite Elem. Anal. Des. 46, 751–759. doi:10.1016/j.finel.2010.04.004

Lee, E.-H., Yang, D.-Y., Yoon, J.W., Yang, W.-H., 2015. Numerical modeling and analysis for forming process of dual-phase 980 steel exposed to infrared local heating. Int. J. Solids Struct. 75, 211–224. doi:10.1016/j.ijsolstr.2015.08.014

Li, D., Ghosh, A., 2003. Tensile deformation behavior of aluminum alloys at warm forming temperatures. Mater. Sci. Eng. A 352, 279–286. doi:10.1016/S0921-5093(02)00915-2

Li, D., Ghosh, A.K., 2004. Biaxial warm forming behavior of aluminum sheet alloys. J. Mater. Process. Technol. 145, 281–293. doi:10.1016/j.jmatprotec.2003.07.003

Logan, R.W., Hosford, W.F., 1980. Upper-bound anisotropic yield locus calculations

assuming ⟨111⟩ -pencil glide. Int. J. Mech. Sci. 22, 419–430. doi:10.1016/0020-7403(80)90011-9

Manach, P.Y., Coër, J., Jégat, A., Laurent, H., Yoon, J.W., 2016. Benchmark 3 - Springback of an Al-Mg alloy in warm forming conditions. J. Phys. Conf. Ser. 734, 22003. doi:10.1088/1742-6596/734/2/022003

Martins, J.M.P., Alves, J.L., Neto, D.M., Oliveira, M.C., Menezes, L.F., 2016. Numerical analysis of different heating systems for warm sheet metal forming. Int. J. Adv. Manuf. Technol. 83, 897–909. doi:10.1007/s00170-015-7618-9

Martins, J.M.P., Neto, D.M., Alves, J.L., Oliveira, M.C., Laurent, H., Andrade-Campos, A., Menezes, L.F., 2017. A new staggered algorithm for thermomechanical coupled problems. Int. J. Solids Struct.

Martins, J.M.P., Neto, D.M., Alves, J.L., Oliveira, M.C., Menezes, L.F., 2016. Numerical modeling of the thermal contact in metal forming processes. Int. J. Adv. Manuf. Technol. 87, 1797–1811. doi:10.1007/s00170-016-8571-y

Masubuchi, K., 1980. Analysis of welded structures : residual stresses, distortion, and their consequences. Pergamon Press.

Mazière, M., Luis, C., Marais, A., Forest, S., Gaspérini, M., 2016. Experimental and numerical analysis of the Lüders phenomenon in simple shear. Int. J. Solids Struct. doi:10.1016/j.ijsolstr.2016.07.026

Menezes, L.F., Neto, D.M., Oliveira, M.C., Alves, J.L., 2011. Improving Computational Performance through HPC Techniques: case study using DD3IMP in-house code. AIP Conf. Proc. 1353, 1220–1225.

Menezes, L.F., Teodosiu, C., 2000. Three-dimensional numerical simulation of the deep-drawing process using solid finite elements. J. Mater. Process. Technol. 97, 100–106.

Nagata, T., 2005. Simple local interpolation of surfaces using normal vectors. Comput. Aided Geom. Des. 22, 327–347.

Naka, T., Torikai, G., Hino, R., Yoshida, F., 2001. The effects of temperature and forming speed on the forming limit diagram for type 5083 aluminum–magnesium alloy sheet. J. Mater. Process. Technol. 113, 648–653. doi:10.1016/S0924-0136(01)00650-1

Neto, D.M., Diogo, C.M.A., Neves, T.F., Oliveira, M.C., Alves, J.L., Menezes, L.F., 2016. Remapping algorithms: application to trimming operations in sheet metal forming. J. Phys. Conf. Ser. 734, 32046. doi:10.1088/1742-6596/734/3/032046

Neto, D.M., Oliveira, M.C., Alves, J.L., Menezes, L.F., 2014a. Influence of the plastic anisotropy modelling in the reverse deep drawing process simulation. Mater. Des. 60, 368–379. doi:10.1016/j.matdes.2014.04.008

Neto, D.M., Oliveira, M.C., Menezes, L.F., Alves, J.L., 2014b. Applying Nagata patches to smooth discretized surfaces used in 3D frictional contact problems. Comput. Methods Appl. Mech. Eng. 271, 296–320.

Neto, D.M., Oliveira, M.C., Menezes, L.F., Alves, J.L., 2013. Nagata patch interpolation using surface normal vectors evaluated from the IGES file. Finite Elem. Anal. Des. 72, 35–46.

Oliveira, M.C., Alves, J.L., Menezes, L.F., 2008. Algorithms and Strategies for Treatment of Large Deformation Frictional Contact in the Numerical Simulation of Deep Drawing

Process. Arch. Comput. Methods Eng. 15, 113–162.

Palumbo, G., Sorgente, D., Tricarico, L., Zhang, S.H., Zheng, W.T., 2007. Numerical and experimental investigations on the effect of the heating strategy and the punch speed on the warm deep drawing of magnesium alloy AZ31. J. Mater. Process. Technol. 191, 342–346. doi:10.1016/j.jmatprotec.2007.03.095

Park, T., Chung, K., 2012. Non-associated flow rule with symmetric stiffness modulus for isotropic-kinematic hardening and its application for earing in circular cup drawing. Int. J. Solids Struct. 49, 3582–3593. doi:10.1016/j.ijsolstr.2012.02.015

Rahmaan, T., Bardelcik, A., Imbert, J., Butcher, C., Worswick, M.J., 2016. Effect of strain rate on flow stress and anisotropy of DP600, TRIP780, and AA5182-O sheet metal alloys. Int. J. Impact Eng. 88, 72–90. doi:10.1016/j.ijimpeng.2015.09.006

Simões, V.M., Oliveira, M.C., Neto, D.M., Cunha, P.M., Laurent, H., Alves, J.L., Menezes, L.F., 2017. Numerical study of springback using the split-ring test: influence of the clearance between the die and the punch. Int. J. Mater. Form.

Takuda, H., Mori, K., Masuda, I., Abe, Y., Matsuo, M., 2002. Finite element simulation of warm deep drawing of aluminium alloy sheet when accounting for heat conduction. J. Mater. Process. Technol. 120, 412–418. doi:10.1016/S0924-0136(01)01180-3

Tardif, N., 2012. Determination of anisotropy and material hardening for aluminum sheet metal. Int. J. Solids Struct. 49, 3496–3506. doi:10.1016/j.ijsolstr.2012.01.011

Toros, S., Ozturk, F., Kacar, I., 2008. Review of warm forming of aluminum–magnesium alloys. J. Mater. Process. Technol. 207, 1–12. doi:10.1016/j.jmatprotec.2008.03.057

van den Boogaard, A.H., Bolt, P.J., Werkhoven, R.J., 2001. Modeling of AlMg Sheet Forming at Elevated Temperatures. Int. J. Form. Process. 4, 361–375. doi:10.3166/ijfp.4.361-375

van den Boogaard, A.H., Huétink, J., 2006. Simulation of aluminium sheet forming at elevated temperatures. Comput. Methods Appl. Mech. Eng. 195, 6691–6709. doi:10.1016/j.cma.2005.05.054

Vaz, M., Lange, M.R., 2016. Thermo-mechanical coupling strategies in elastic–plastic problems. Contin. Mech. Thermodyn. 1–11. doi:10.1007/s00161-016-0537-7

Wilson, D.V., 1988. Aluminium versus steel in the family car — the formability factor. J. Mech. Work. Technol. 16, 257–277. doi:10.1016/0378-3804(88)90055-1

Wriggers, P., Miehe, C., Kleiber, M., Simo, J.C., 1992. On the coupled thermomechanical treatment of necking problems via finite element methods. Int. J. Numer. Methods Eng. 33, 869–883. doi:10.1002/nme.1620330413

Yilmaz, A., 2011. The Portevin–Le Chatelier effect: a review of experimental findings. Sci. Technol. Adv. Mater. 12, 63001. doi:10.1088/1468-6996/12/6/063001

Yoon, J.W., Dick, R.E., Barlat, F., 2011. A new analytical theory for earing generated from anisotropic plasticity. Int. J. Plast. 27, 1165–1184. doi:10.1016/j.ijplas.2011.01.002

Yoon, J.W., Yang, D.Y., Chung, K., Barlat, F., 1999. A general elasto-plastic finite element formulation based on incremental deformation theory for planar anisotropy and its application to sheet metal forming. Int. J. Plast. 15, 35–67. doi:10.1016/S0749-6419(98)00059-X

Zhang, Y., Zhu, P., Chen, G., 2007. Lightweight Design of Automotive Front Side Rail Based on Robust Optimisation. Thin-Walled Struct. 45, 670–676. doi:10.1016/j.tws.2007.05.007

Zhao, K., Wang, B., Chang, Y., Tang, X., Yan, J., 2015. Comparison of the methods for calculating the interfacial heat transfer coefficient in hot stamping. Appl. Therm. Eng. 79, 17–26. doi:10.1016/j.applthermaleng.2015.01.018